\documentclass[lettersize,journal]{IEEEtran}
\usepackage{amsmath,amsfonts}
\usepackage{algorithmic}
\usepackage{algorithm}
\usepackage{array}
\usepackage{amsmath, amssymb}
\usepackage{bm}
\usepackage{tabularray}
\usepackage{mathrsfs}
\usepackage{MnSymbol} 
\usepackage[mathcal]{eucal}
\usepackage[caption=false,font=normalsize,labelfont=sf,textfont=sf]{subfig}
\usepackage{textcomp}
\usepackage{stfloats}
\usepackage{booktabs}
\usepackage{url}
\usepackage{verbatim}
\usepackage{threeparttable}
\usepackage{multirow}
\usepackage{graphicx}
\usepackage{etoolbox}
\usepackage{cite}
\begin{document}

\title{SPAC: Sampling-based Progressive Attribute Compression  for Dense Point Clouds}

\author{Xiaolong Mao, Hui Yuan~\IEEEmembership{Senior Member,~IEEE}, Tian Guo, Shiqi Jiang, Raouf Hamzaoui~\IEEEmembership{Senior Member,~IEEE}, and Sam Kwong~\IEEEmembership{Fellow,~IEEE}
        % <-this % stops a space
\thanks{This work was supported in part by the National Natural Science Foundation of China under Grants 62222110 and 62172259, the High-end Foreign Experts Recruitment Plan of Chinese Ministry of Science and Technology under Grant G2023150003L, the Taishan Scholar Project of Shandong Province (tsqn202103001), the Natural Science Foundation of Shandong Province under Grant ZR2022ZD38, and the OPPO research fund. (Corresponding author: Hui Yuan)

Xiaolong Mao Hui Yuan,  Tian Guo, and Shiqi Jiang are with the School of Control Science and Engineering, Shandong University, Jinan 250061, China (e-mail: huiyuan@sdu.edu.cn).

Raouf Hamzaoui is with the School of Engineering and Sustainable
Development, De Montfort University, LE1 9BH Leicester, U.K. (e-mail: rhamzaoui@dmu.ac.uk).

Sam Kwong is with the School of Data Science, Lingnan University, Hong Kong (e-mail: samkwong@ln.edu.hk).}% <-this % stops a space        
}

\markboth{Journal of \LaTeX\ Class Files,~Vol.~14, No.~8, August~2021}%
{Shell \MakeLowercase{\textit{\textit{et al.}}}: A Sample Article Using IEEEtran.cls for IEEE Journals}


\IEEEoverridecommandlockouts
\maketitle

\begin{abstract}
We propose an end-to-end attribute compression method for dense point clouds. The proposed method combines a frequency sampling module, an adaptive scale feature extraction module with geometry assistance, and a global hyperprior entropy model. The frequency sampling module uses a Hamming window and the Fast Fourier Transform to extract high-frequency components of the point cloud. The difference between the original point cloud and the sampled point cloud is divided into multiple sub-point clouds. These sub-point clouds are  then partitioned using an octree, providing a structured input for feature extraction. The feature extraction module integrates adaptive convolutional layers and uses offset-attention to capture both local and global features. Then, a geometry-assisted attribute feature refinement module is used to refine the extracted attribute features. Finally, a global hyperprior model is introduced for entropy encoding. This model propagates hyperprior parameters from the deepest (base) layer to the other layers, further enhancing the encoding efficiency. At the decoder, a mirrored network is used to progressively restore features and reconstruct the color attribute through transposed convolutional layers. The proposed method encodes base layer information at a low bitrate and progressively adds enhancement layer information to improve reconstruction accuracy. Compared to the latest G-PCC test model (TMC13v23) under the MPEG common test conditions (CTCs), the proposed method achieved an average Bj{\o}ntegaard delta bitrate reduction of 24.58\% for the Y component (21.23\% for YUV combined) on the MPEG Category Solid dataset and 22.48\% for the Y component (17.19\% for YUV combined) on the MPEG Category Dense dataset. This is the first instance of a learning-based codec outperforming the G-PCC standard on these datasets under the MPEG CTCs. Our source code will be made publicly available on https://github.com/sduxlmao/SPAC.
\end{abstract}


\section{Introduction}
\IEEEPARstart{W}{ith} the rapid advancement of three-dimensional(3D) acquisition technologies, point clouds have gained popularity as a medium for representing real-world scenes and objects in applications such as augmented/virtual reality (AR/VR)\cite{ref1}, immersive communication \cite{ref2,ref3}, cultural heritage preservation\cite{ref4}, and autonomous driving\cite{ref5}. Typically, a point cloud comprises millions of unordered points across a 3D surface. Each point is characterized by geometric information, such as its Cartesian coordinates \textit{(x, y, z)}, and attributes like color, normals, and reflectance\cite{ref6}. A small colored static point cloud with one million points, representing each point's geometry with 10-bit precision and RGB color with 8-bit precision, requires approximately 54MB of data. The transmission of dynamic point clouds imposes additional bandwidth demands. For a dynamic point cloud with 25 frames per second, the bitrate can reach 1.35GB/s. Without efficient compression, meeting such requirements would be infeasible for existing networks.

Given the inherent unordered and sparse nature of point clouds, significant research efforts over the past few decades have focused on efficiently exploiting spatial neighbourhood organization to explore spatial correlations through methods such as 1D traversal, 2D projection, or 3D spatial indexing. Based on these past studies, the Moving Picture Experts Group (MPEG) developed two standards for 3D point cloud compression: video-based point cloud compression (V-PCC) and geometry-based point cloud compression (G-PCC). V-PCC\cite{ref7} projects dynamic 3D point clouds onto 2D planes, enabling compression using existing video coding standards, such as H.265/HEVC\cite{ref8} and H.266/VVC\cite{ref9}. G-PCC \cite{ref10} uses geometric decomposition to capitalize on the 3D correlations present in both static and dynamically acquired point clouds. G-PCC focuses on directly processing the geometry, aiming to preserve the spatial integrity and characteristics of the original 3D structure. 

Beyond traditional point cloud compression methods, deep learning techniques have emerged as a promising approach. This novel approach exploits neural networks to further enhance the efficiency of point cloud processing. Deep learning-based point cloud compression methods primarily focus on learning compact representations of point clouds by training neural networks to explore the intricate spatial relationships and patterns. These methods typically involve an encoder-decoder architecture. The encoder transforms the input point cloud into a lower-dimensional, compact representation. The decoder reconstructs the point cloud from the compact representation, aiming to retain as much of the original detail and structure as possible. Recognizing the potential of deep learning, MPEG is planning to release a call for proposal for artificial intelligence-based point cloud compression \cite{ref11}. 

Building on the fundamental concepts of point cloud compression, especially the hierarchical octree structure, we propose a novel approach that uses the spatial hierarchical structure of point clouds for deep learning-based progressive point cloud attribute compression. Specifically, the contributions of this paper are as follows.

\begin{itemize}

\item We propose an end-to-end multi-layer attribute coding method for dense point clouds. Our method features a frequency-based sampling module, an adaptive scale feature extraction module, a geometry-assisted attribute feature refinement module, and a global hyperprior entropy model.

\item The frequency-based sampling module uses Hamming window-based pre-processing and the Fast Fourier Transform (FFT) to progressively extract high-frequency components, where color attributes exhibit significant variations.

\item The adaptive scale feature extraction module uses deeper networks for layers with high-frequency components and a lightweight network for layers with low-frequency components.

\item We propose a geometry-assisted attribute feature refinement module to enhance the learning of local variations by using point normals.

\item We introduce the hyperprior entropy model in the deepest layer (corresponding to the highest-frequency contour regions) to guide the encoding of all other layers. Simultaneously, we incorporate adaptive quantization into the hyperprior entropy model, which adaptively adjusts quantization parameters based on the characteristics of the point cloud, further reducing the bitrate.

\item The proposed method encodes and transmits the base layer at a low bitrate and progressively adds enhancement layers at higher bitrates to improve reconstruction quality. This allows for dynamic adjustment of the bitrate according to the available bandwidth.

\item Our method is the first end-to-end deep learning method to outperform G-PCC TMC13v23 under the MPEG common test conditions (CTCs). The Bj{\o}ntegaard delta bitrate(BD-BR) reduction and BD-peak signal-to-noise ratio(PSNR) increase are significant, reaching up to 23.53\% and 0.66 dB, respectively.

\end{itemize}

The rest of this paper is organized as follows. In Section II, we briefly review related work on traditional and deep learning-based point cloud compression. In Section III, we provide a theoretical motivation for the proposed method and formulate the problem mathematically. In Section IV, building on Section III, we present the proposed method. In section V, we compare our method to the latest G-PCC test model and to two state-of-the-art deep learning based methods. Finally,we provide conclusions and suggest future work in Section VI.

\section{Related Work}
We classify point cloud compression methods into three categories: MPEG G-PCC, traditional hybrid methods, and deep learning-based methods.

\subsection{MPEG G-PCC}

In G-PCC\cite{ref12}, the geometry encoding process begins by normalizing the spatial coordinates of the point cloud to ensure that the coordinates fall within a fixed range, thus preparing the data for subsequent voxelization and encoding. The voxelization step partitions the continuous three-dimensional space into equal-sized voxels, each containing one or more points, thereby discretizing the point cloud and facilitating subsequent encoding process. G-PCC employs three geometric encoding methods \cite{ref13}. Octree encoding leverages the octree structure to encode the voxelized point cloud by recursively subdividing the three-dimensional space into eight subspaces, each further subdivided based on whether it contains points. This method achieves efficient spatial partitioning and hierarchical representation, making it suitable for point clouds with a clear spatial hierarchy. Trisoup encoding, on the other hand, uses triangulation to represent the point cloud as a triangular mesh for efficient compression. This method is highly efficient in representing planar and smooth surfaces, and is usually used together with octree. Predictive tree encoding uses spatial prediction methods to predict the position of each point based on its neighboring points by constructing a prediction tree and using the information from already encoded points. This method is very efficient to exploit the spatial correlation in LiDAR point clouds. The final output of the geometry encoding process is a geometry bitstream generated through arithmetic encoding.

In the attribute encoding process, the attribute transfer step is first conducted to ensure that attribute information is accurately transferred from the original point cloud to the voxelized and geometry reconsturcted point cloud. Subsequently, G-PCC can select three attribute encoding methods, level of detail (LOD)-based predictive transform (PT) or lifting transform (LT), and region adaptive hierarchical transform (RAHT),  to efficiently compress the point cloud aiming at different applications. PT encodes attributes by predicting the current point’s attribute based on its neighboring points, leveraging the correlation between attributes to improve compression efficiency. To further improve the efficiency, Guo \textit{et al.} \cite{ref6} proposed a dependence-based coarse-to-fine approach for distortion accumulation in PT. LT decomposes and reconstructs the attribute information recursively into a more compact representation, effectively reducing data redundancy and enhancing encoding efficiency. RAHT encodes the attributes by sequentially decomposing and transforming the attribute information into 3D Haar wavelet coefficients, which is more efficient for dense point clouds \cite{ref14}. The final attribute bitstream is generated through quantization and arithmetic encoding. Overall, these methods enable G-PCC to efficiently compress point clouds to satisfy different practical applications.
\iffalse
\begin{figure}[!t]
\setlength{\abovecaptionskip}{-0.1cm}   %调整图片标题与图距离
\setlength{\belowcaptionskip}{-3cm}   %调整图片标题与下文距离
\centering
\includegraphics[width=2in]{GPCC.png}
\caption{G-PCC encoder diagram.}
\label{fig_1}
\end{figure}
\fi
\subsection{Traditional Hybrid Framework-based Compression}

\textbf{Geometry Compression:} Drawing inspiration from the efficient spatial partitioning and hierarchical representation of G-PCC, Huang \textit{et al.} \cite{ref15} implemented progressive geometric encoding using an octree structure. In \cite{ref16}, octree encoding was combined with graph Fourier transform (GFT) for hybrid lossy geometric encoding. Additionally, Schnabel and Klein \cite{ref17} employed surface-based moving least squares  approximation in their breadth-first partitioning to predict occupancy codes and therefore achieving octree encoding.

Besides octree structures, the KD-tree (k-dimensional binary tree) \cite{ref18} is another way to ensure real-time tracking and registration of related points in high-density point clouds. It efficiently manages spatial data, enabling rapid nearest-neighbor search and other spatial queries, which are critical in many point cloud applications. Kathariya \textit{et al.} \cite{ref19} proposed a point cloud geometry coding method that combines binary and quadtree structures. The method divides point cloud data into smaller blocks using a binary tree structure, encoding planar surfaces with a quadtree and non-planar surfaces with an octree. This approach maintains geometric details while improving compression efficiency and supports applications with different resolution requirements.

\textbf{Attribute Compression:} In addition to the attribute encoding methods of MPEG G-PCC, GFT and its variants are common examples of transform-based attribute encoding methods, as studied in \cite{ref20}. Researchers such as Zhang \textit{et al.} \cite{ref22} directly used GFT, while others like Robert \textit{et al.} \cite{ref24} and Song \textit{et al.} \cite{ref25} proposed methods based on block prediction and GFT. Cohen \textit{et al.} \cite{ref26} and Ricardo \textit{et al.} \cite{ref27} introduced approaches using 3D block prediction and hierarchical transforms, respectively. Chen \textit{et al.} \cite{ref28} developed a self-loop weighted graph using normalized graph Laplacian to define GFT. Shao \textit{et al.} \cite{ref29} introduced a new binary tree-based point cloud content partition and explored GFT with optimized Laplacian sparsity, to achieve better energy compaction and compression efficiency. Gu \textit{et al.} \cite{ref30} proposed an effective compression scheme for the attributes of voxelized 3D point clouds. Additionally, Gaussian process transform explored in \cite{ref21}, uses the Karhunen–Lo\`eve transform matrix of a Gaussian process to transform a point cloud into compact spectrum representation. Liu \textit{et al.} \cite{ref23} propose a sparse representation strategy based on virtual adaptive sampling to remove redundancy among points, where the bases of graph transform and discrete cosine transform are used as candidates for the complete dictionary.  Li \textit{et al.} \cite{ref68} proposed a novel p-Laplacian embedding graph dictionary learning framework for efficient 3D point cloud attribute compression, leveraging signal statistics and high-order geometric structures. However, these transform-based compression methods for point cloud attributes rely on prior assumptions about the point cloud data, limiting their universality and flexibility.

\subsection{Deep Learning-based Compression}

Due to the significant advantages of neural networks, many learning-based point cloud compression methods have been proposed. One key advantage of the learning methods is their capacity to adaptively learn from data, enabling more efficient handling of the variability and complexity inherent in point clouds.

\textbf{Geometry Compression:} In early research, Huang \textit{et al.} \cite{ref31} used neural networks to design encoders and decoders, extracting features from the original models, followed by further compression of codewords using sparse coding. Subsequently, an increasing number of studies emerged. By integrating a learned hierarchical feature extraction and encoding process, PCGCv2 \cite{ref32} significantly outperforms traditional methods in terms of compression efficiency. It reconstructs geometry details by pruning false voxels and extracting true occupied voxels using binary classification, which follows a hierarchical, coarse-to-fine refinement. \iffalse Wang \textit{et al.} \cite{ref33} applied a VAE based on stacked deep neural networks (DNNs) to efficiently compress point cloud geometry. In their subsequent research \cite{ref34}, they proposed a neural network based on sparse convolutions,known as SparsePCGC, to capture and embed spatial correlations.\fi
 Nguyen \textit{et al.}  \cite{ref35} used a multi-scale architecture to model voxel occupancy in an order from coarse to fine. Zhang \textit{et al.} \cite{ref36} proposed a method called YOGA, which enables variable-rate encoding with a lightweight single neural model. The method leverages sparse convolutions and parallel entropy coding. Wang \textit{et al.} \cite{ref37} demonstrated the potential of combining octree structures with neural networks for compression and proposed OctSqueeze. Que \textit{et al.} \cite{ref38} introduced VoxelContext-Net, which is capable of compressing both static and dynamic point cloud geometries. The method uses voxel context to efficiently compress octree-structured data. Building upon \cite{ref37} and \cite{ref38}, Fu \textit{et al.} \cite{ref39} proposed a multi-context deep learning framework named OctAttention, to encode octree symbol sequences losslessly by gathering information from neighbouring and ancestor nodes. Sun \textit{et al.} \cite{ref40} proposed a general structure that enhances existing context models by introducing context feature residuals and a multi-layer perceptron branch. The structure improves the accuracy of node occupancy probability distribution prediction and gradient updates in point cloud geometry compression. The authors \cite{ref41} also proposed an attention-based child node number prediction module to further enhance the context models. Akhtar \textit{et al.} \cite{ref42} proposed a lossy geometry compression scheme that predicts the latent representation of the current frame using the previous frame by employing a novel feature space inter-prediction network. Wu \textit{et al.} \cite{ref69} introduced a method to improve human point cloud compression by using a geometric prior, enhancing performance while preserving quality.

\textbf{Attribute Compression:} Neural networks in point cloud attribute compression are emerging, predominantly in two approaches. The first approach projects irregular structures onto regular ones, facilitating the processing of irregular point cloud inputs. For example, Quach \textit{et al.} \cite{ref43} trained a lossy folding network to map 3D attributes onto 2D grids, compressing them with video codecs. The second approach uses autoencoder frameworks with 3D dense convolutions for attribute compression. The first end-to-end point cloud attribute compression framework, inspired by PointNet \cite{ref44}, was presented in \cite{ref45}, which encodes and decodes attributes directly without voxelization or point projection. With the aid of geometric information, it leverages the spatial correlations among points and the nonlinear relationships between attribute features. This adaptability results in better preservation of details and more efficient compression, particularly for complex or irregular structures that traditional methods might struggle to compress efficiently. Wang \textit{et al.} \cite{ref46} used a sparse tensor to build a Variational Autoencoder (VAE) for color attribute compression. Pinheiro \textit{et al.} \cite{ref47} introduced a novel lossy attribute compression network based on normalizing flows, named NF-PCAC, capable of achieving performance comparable to G-PCC TMC13v14 encoding. Nguyen \textit{et al.} \cite{ref48} explored the effectiveness of 3D sparse convolutions in lossless attribute compression. Fang \textit{et al.} \cite{ref49} converted point cloud attributes into transformed coefficients and used a compression network, namely 3DAC, to further compress these coefficients into an attribute bitstream. Zhang \textit{et al.} \cite{ref50} proposed a method called ScalablePCAC for scalable point cloud attribute compression. It uses G-PCC at the base layer to encode a downscaled thumbnail point cloud and a learning-based model with symmetric structure at the enhancement layer to compress and restore the full-resolution point cloud. Rudolph \textit{et al.} \cite{ref70} proposed a method to progressively encode point cloud attributes in a learned manner by compressing quantization residuals in the entropy bottleneck. Many methods focus on enhancing the attributes of the encoded point cloud, thus improving the coding efficiency of point cloud compression systems. Ding \textit{et al.} \cite{ref51} proposed a learning-based adaptive loop filter to reduce compression artifacts in point cloud data. Xing \textit{et al.} \cite{ref52,ref53} proposed methods to enhance point cloud color quality: one uses a small-scale U-Net architecture, and the other uses a graph-based convolutional network, both aiming to improve the visual fidelity of point clouds.

In summary, despite technological advancements, the efficiency of these lossy compression methods still falls short when compared to the latest G-PCC standard. Geometry provides a reference structure that enables attributes to be accurately mapped to their correct spatial positions, thereby ensuring high-fidelity reconstruction. Therefore, the MPEG AI- 3D Graphics Compression (3DGC) group recommended to prioritize lossless or near lossless geometry compression \cite{ref54}. Based on this recommendation, we assume that the geometric information of the point cloud is losslessly compressed, and our work primarily focuses on attribute compression  aiming at exceeding the latest G-PCC standard in terms of  rate-distortion (RD) performance.

\section{Problem Formulation}
Sparse signals are well-suited for compression \cite{ref55}.  Therefore, decomposing a complex signal, such as a point cloud, into several sparse parts can achieve high compression efficiency. Moreover, the human visual system (HVS) exhibits different sensitivities at different frequencies. In particular, human eyes are less sensitive to high-frequency details than to low-frequency energy \cite{ref56}. By using frequency decomposition, perceptual redundancy can be exploited during compression to achieve higher efficiency. 

Let $\bm{\mathcal{P}} = (\mathbf{G}, \mathbf{C}) = \{\bm{p}_i = (\mathbf{G}_i, \mathbf{C}_i)\}_{i=1}^M$ represent the point cloud to be compressed, where $\mathbf{G}_i$ and $\mathbf{C}_i$ denote the spatial coordinates and color attribute of point $\bm{p}_i$, respectively. The goal is to decompose $\bm{\mathcal{P}}$ into low-frequency and high-frequency components. To analyse the frequency characteristics of the input point cloud, we use the FFT. In the context of point clouds, the color attributes can be treated as 1D discrete signals based on the input point order. To minimize discontinuities and reduce the leakage effect when applying the FFT, we first use a window function

\begin{equation}
\small
\label{eq1}
\bm{\omega}(n) = 0.54 - 0.46 \cos \left( \frac{2 \pi n}{N-1} \right), \quad n = 0, 1, \ldots, N-1,
\end{equation}

\noindent and obtain

\begin{equation}
\label{eq2}
\mathbb{F}(\bm{\mathbf{C}}) = \text{FFT}(\bm{\omega} \cdot \mathbf{C}),
\end{equation}
\noindent where ``$\cdot$'' denotes dot product, $\mathbb{F}(\bm{\mathbf{C}})$ represents the frequency domain representation of the color attribute. The high-frequency component $\bm{\mathcal{P}}_{\textit{\text{high}}}$ can be obtained by mapping the color attribute of the initial point cloud onto the non-zero positions of the inverse FFT (IFFT) of the high-frequency coefficients in $\mathbb{F}(\bm{\mathbf{C}})$(see Fig. 2 in Section IV):

\begin{equation}
\label{eq3}
\bm{\mathbf{C}}'_{\textit{\text{high}}} = \text{IFFT}(\mathbb{F}(\bm{\mathbf{C}})_{\textit{\text{high}}}),
\end{equation}
\begin{equation}
\label{eq22}
\bm{\mathcal{P}}_{\textit{\text{high}}} = \textit{\text{map}}(\bm{\mathcal{P}}, \bm{\mathbf{C}}'_{\textit{\text{high}}}) = (\mathbf{G}_{\textit{\text{high}}}, \mathbf{C}_{\textit{\text{high}}}).
\end{equation}

\noindent where $\mathbb{F}(\bm{\mathbf{C}})_{\textit{\text{high}}}$ represents the high-frequency coefficients. The low-frequency component $\bm{\mathcal{P}}_{\textit{\text{low}}}$ can then be regarded as the difference set (residual) between the points in the original point cloud and $\bm{\mathcal{P}}_{\textit{\text{high}}}$:

\begin{figure*}[!t]
\setlength{\abovecaptionskip}{-0.2cm}
\centering
\includegraphics[width=6.5in]{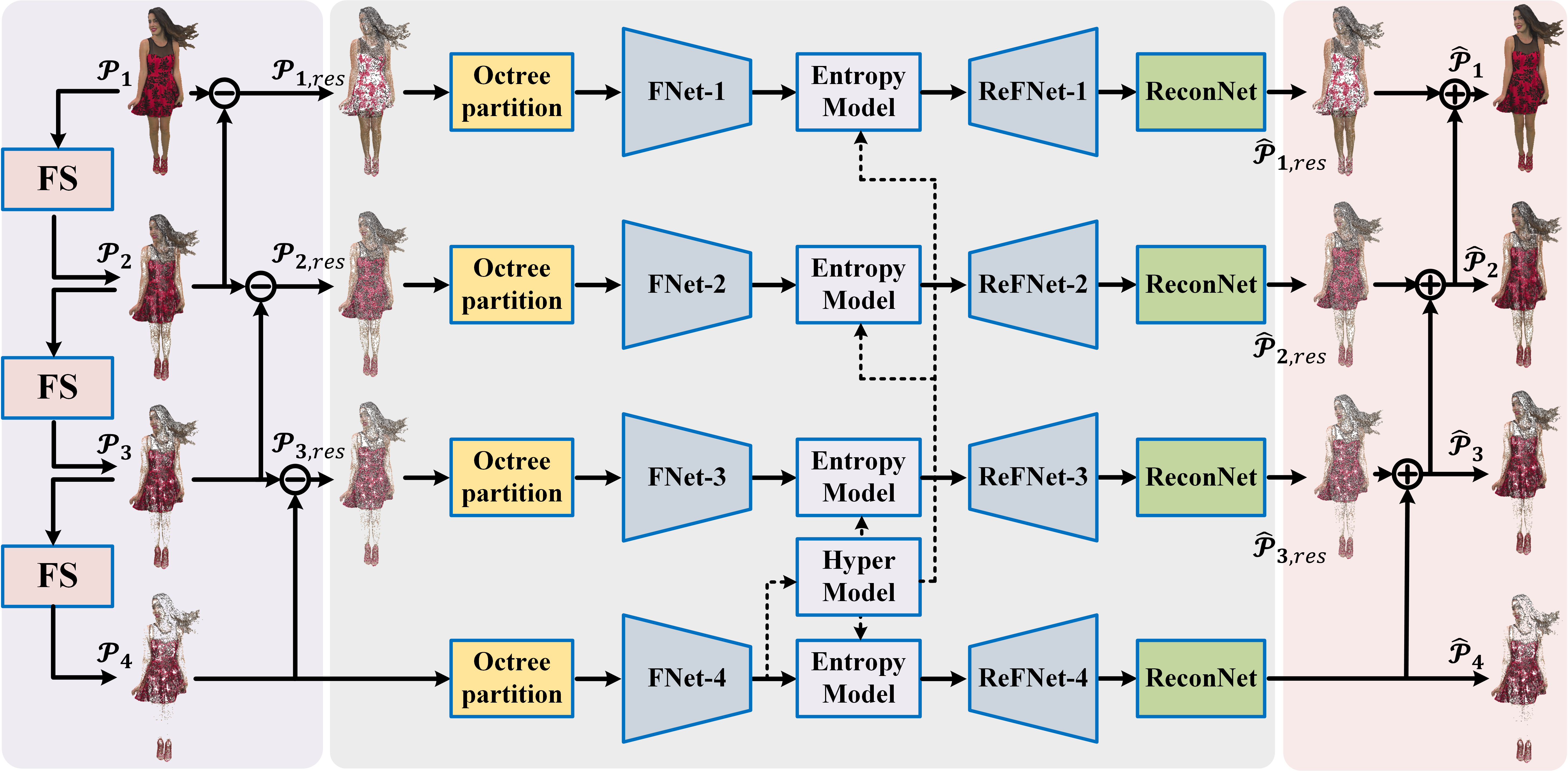}
\caption{SPAC architecture. The input point cloud is processed through a Frequency Sampling (FS) module, resulting in sampled point clouds $\bm{\mathcal{P}}_1, \bm{\mathcal{P}}_2, \bm{\mathcal{P}}_3, \bm{\mathcal{P}}_4$. Residual point clouds $\bm{\mathcal{P}}_{1,\text{res}}, \bm{\mathcal{P}}_{2,\text{res}}, \bm{\mathcal{P}}_{3,\text{res}}$ are then obtained through set difference. The residual point clouds are further partitioned using octree and fed into Feature Extraction Networks (FNet-1, FNet-2, FNet-3, FNet-4) for feature extraction. Next, the extracted features are encoded using entropy coding, where the entropy encoder of the deepest layer incorporates a hyperprior entropy model to enhance the encoding efficiency of all layers. During decoding, the features are processed by the corresponding decoding networks (ReFNet-1, ReFNet-2, ReFNet-3, ReFNet-4), and the decoded residual point clouds are reconstructed through the Reconstruction Network (ReconNet). The reconstructed residual point clouds are progressively concatenated with higher-level point clouds, ultimately resulting in the final reconstructed point clouds $\hat{\bm{\mathcal{P}}}_1, \hat{\bm{\mathcal{P}}}_2, \hat{\bm{\mathcal{P}}}_3, \hat{\bm{\mathcal{P}}}_4$.
}
\label{fig_2}
\end{figure*}

\begin{equation}
\label{eq4}
\bm{\mathcal{P}}_{\textit{\text{low}}} = \bm{\mathcal{P}} \ominus \bm{\mathcal{P}}_{\textit{\text{high}}} = (\mathbf{G}_{\textit{\text{low}}}, \mathbf{C}_{\textit{\text{low}}}),
\end{equation}
\noindent where $\ominus$ denotes the set difference operation.
\iffalse 
Therefore, the frequency-based sampling  for color attribute that we focused in this paper can be represented as
\begin{equation}
\label{eq5}
( \mathbf{C}_{\textit{\text{high}}}) = \textit{\text{Sampling}}_{\textit{\text{high}}}( \mathbf{C}),
\end{equation}
\begin{equation}
\label{eq6}
( \mathbf{C}_{\textit{\text{low}}}) = \mathbf{C} \ominus \mathbf{C}_{\textit{\text{high}}}.
\end{equation}
\fi 

The corresponding multi-layer progressive encoding-decoding structure is then designed to handle different frequency components separately, providing an efficient hierarchical representation. Each layer focuses on encoding and decoding specific frequency components, progressively refining the reconstruction.

Let $L$ denote the number of layers in the encoding-decoding hierarchy. For each layer $l \in \{1, \ldots, L\}$, we define $\bm{\mathcal{P}}_l$ as the input point cloud, $\bm{R}_l$ as the encoded bitstream, and $\hat{\bm{\mathcal{P}}}_l$ as the reconstructed point cloud. The encoding function $\mathbb{E}_l$ and decoding function $\mathbb{D}_l$ for each layer are represented as

\begin{equation}
\label{eq7}
\bm{R}_l = \mathbb{E}_l(\bm{\mathcal{P}}_{l,\text{res}} \mid \bm{\Theta}_{\mathbb{E},l}),
\end{equation}
\begin{equation}
\label{eq8}
\hat{\bm{\mathcal{P}}}_l = \mathbb{D}_l(\bm{R}_l \mid \bm{\Theta}_{\mathbb{D},l}),
\end{equation}

\noindent where $\bm{\Theta}_{\mathbb{E},l}$ and $\bm{\Theta}_{\mathbb{D},l}$ are the network parameters of the encoder and decoder in the $l$-th layer, and the input of $\mathbb{E}_l$ is the residual between $\bm{\mathcal{P}}_l$ and $\bm{\mathcal{P}}_{l+1}$:

\begin{equation}
\label{eq9}
\bm{\mathcal{P}}_{l,\text{res}} = \bm{\mathcal{P}}_l \ominus \bm{\mathcal{P}}_{l+1},
\end{equation}

\noindent which corresponds to $(\mathbf{G}_{\textit{\text{low}}}, \mathbf{C}_{\textit{\text{low}}})$ of layer $l$.

The hierarchical nature of the proposed encoding and decoding method inherently supports scalability. At low bitrates, only the base layer is encoded and transmitted, providing a coarse reconstruction of the point cloud.  Additional layers ($l < L$) can be progressively encoded, improving reconstruction quality while increasing the bitrate. Combining the frequency-based sampling and the multi-layer progressive encoding and decoding paradigm, the overall objective is to minimize the distortion $D(\bm{\mathcal{P}}, \hat{\bm{\mathcal{P}}})$ subject to a bitrate constraint:
\begin{equation}
\label{eq10}
\min_{\{\bm{\Theta}_{\mathbb{E},l}, \bm{\Theta}_{\mathbb{D},l}\}_{l=1}^{L}} D(\bm{\mathcal{P}}, \hat{\bm{\mathcal{P}}}) \quad \text{subject to} \quad \sum_{l=1}^{L} R\left( \mathbb{E}_l(\bm{\mathcal{P}}_{l,\text{res}}) \right) \leq R_c,
\end{equation}

\noindent where $R_c$ is the bitrate constraint. To  solve the constrained optimization problem (9), we introduce a Lagrange multiplier $\lambda$ to transform it into the  unconstrained optimization problem \cite{ref57}:
\begin{equation}
\label{eq11}
\min_{\{\bm{\Theta}_{\mathbb{E},l}, \bm{\Theta}_{\mathbb{D},l}\}_{l=1}^{L}} \left[ D(\bm{\mathcal{P}}, \hat{\bm{\mathcal{P}}}) + \lambda \sum_{l=1}^{L} R\left( \mathbb{E}_l(\bm{\mathcal{P}}_{l,\text{res}}) \right) \right].
\end{equation}

\section{Proposed Method}
The proposed sampling-based progressive attribute compression method (SPAC) implements the concept of layered transformation through frequency sampling (Fig. 1). We denote the original point cloud $\bm{\mathcal{P}}_1 = \{ (\mathbf{G}_1, \mathbf{C}_1), \ \mathbf{G}_1 \in \mathbb{R}^{M_1 \times 3}, \mathbf{C}_1 \in \mathbb{R}^{M_1 \times 3} \}$, where $M_1$ is the number of points in the original point cloud, $\mathbf{G}_1$ represents the 3D coordinates of the points, and $\mathbf{C}_1$ denotes the associated color attributes. A key component of SPAC is the proposed frequency sampling (FS) module which samples components where the attribute varies significantly. This results in subsampled point clouds $\bm{\mathcal{P}}_2 \in \mathbb{R}^{M_2 \times 6}, \bm{\mathcal{P}}_3 \in \mathbb{R}^{M_3 \times 6}, \bm{\mathcal{P}}_4 \in \mathbb{R}^{M_4 \times 6}$, etc., undergoing adaptive octree partitioning to ensure efficient compression. Additionally, the network uses adaptive scale feature extraction with different depth at each layer, adapting attribute redundancy across different frequencies. As a consequence, the decoder uses a coarse-to-fine reconstruction strategy, mirroring the encoding modules to effectively decode the point cloud attributes.

\begin{figure}[!t]
\centering
\includegraphics[width=3.5in]{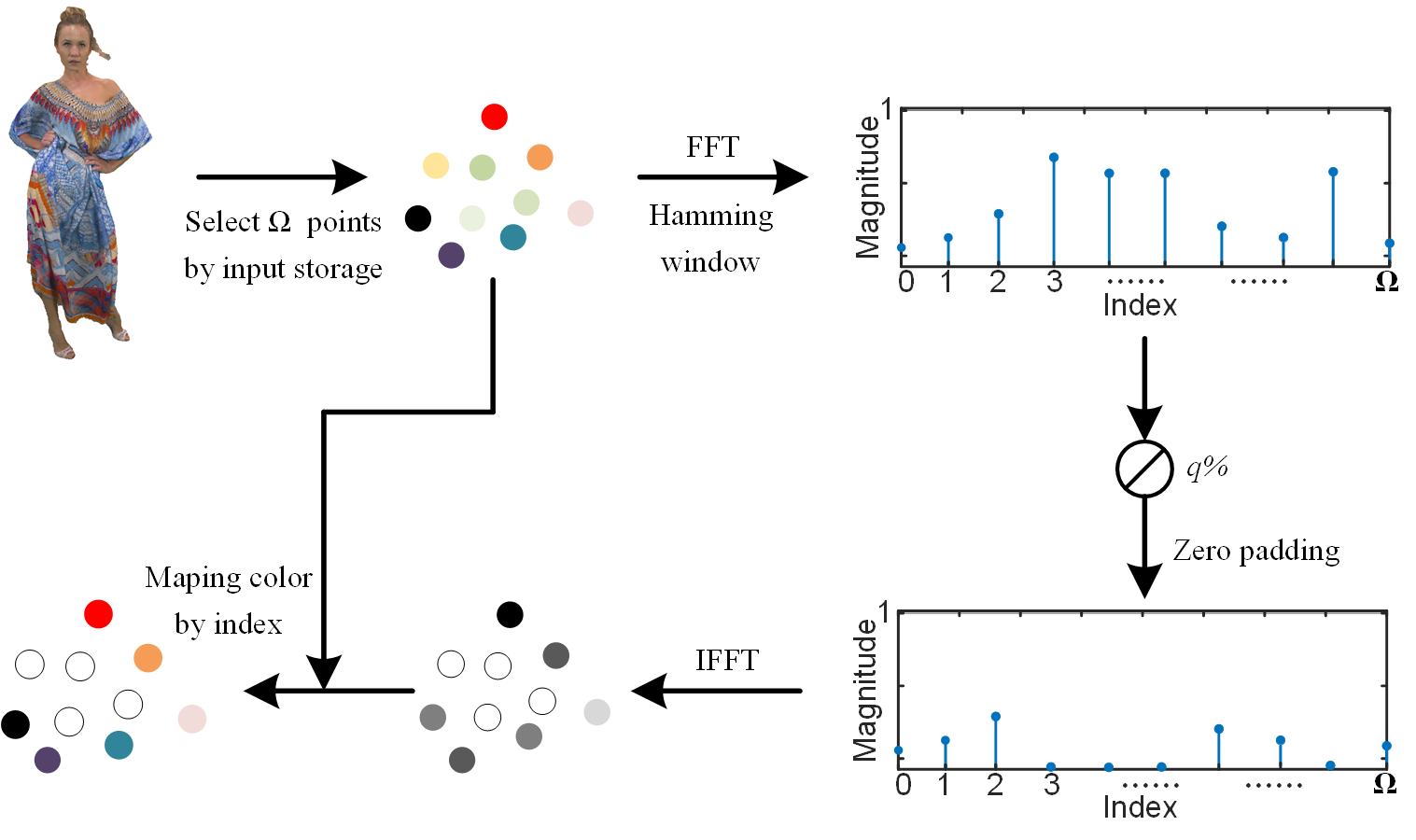}
\caption{Structure of the FS module. Each \textbf{$\Omega$} points in a group are processed using a Hamming window and the FFT. The coefficients whose magnitude is smaller than or equal to \textit{q}\% of the largest magnitude are retained, while the other coefficients are set to zero. Afterward, the IFFT is used to transform the processed coefficients back to the spatial domain. The input \textbf{$\Omega$} points in the group are then mapped to the non-zero positions  (as shown in the black and gray dots in the figure) of the IFFT results to obtain the high-frequency component of this group.}
\label{fig_3}
\end{figure}

\subsection{FS Module}
As depicted in Fig. 2, the FS module is designed to selectively sample points from the input point cloud, effectively preserving high-frequency components. This module first divides the input points of $\bm{\mathcal{P}}_1$ into several groups, each containing \textbf{$\Omega$} points. To distinguish between high and low-frequency components of a group, we apply the FFT to transform the color attributes into the frequency domain. Prior to this transformation, as mentioned in Section III, the input points are pre-processed by using a Hamming window, to mitigate spectral leakage:
\begin{equation}
\label{eq13}
\mathbf{C}_{\textit{\text{win}}} = \bm{\omega} \cdot \mathbf{C}_i.
\end{equation}

\noindent Applying the FFT to $\mathbf{C}_{\textit{\text{win}}}$ gives:
\begin{equation}
\label{eq15}
\mathbb{F}(\mathbf{C}_{\textit{\text{win}}})_k = \sum_{m=0}^{\Omega-1} \mathbf{C}_{\textit{\text{win},m}} e^{- \frac{2\pi jkm}{\Omega}},
\end{equation}

\noindent where $k$ is the index of the frequency component, and $j$ is the imaginary unit.

To sample out the high frequency components, in the frequency-domain, we retain the coefficients whose magnitude is smaller than or equal to \textit{q}\% (here we choose \textit{q} as 60) of the maximum magnitude and set the remaining coefficients to zero:

\begin{equation}
\mathbb{F}(\mathbf{C}_{\text{zp}})_k = 
\begin{cases} 
\mathbb{F}(\mathbf{C}_{\text{win}})_k & \text{if magnitude} \leq q\% \times \max(\text{magnitude}) \\
0 & \text{if magnitude} > q\% \times \max(\text{magnitude}).
\end{cases}
\end{equation}

Then, we apply the IFFT on the zero-padded frequency-domain signal $\mathbb{F}(\mathbf{C}_{\text{zp}})$, converting it back to the spatial domain,
\begin{equation}
\label{eq22}
\mathbf{C}'_{\textit{\text{high}}} = \mathbb{F}^{-1}(\mathbf{C}_{\text{zp}}),
\end{equation}
and then map the input point cloud to the sampled one, as shown in Fig. 3,
\begin{equation}
\label{eq22}
\mathbf{C}_{\textit{\text{high}}} = \textit{\text{map}}(\mathbf{C}_{\textit{\text{orig}}}, \bm{C}'_{\textit{\text{high}}}).
\end{equation}

\begin{figure}[!t]
\setlength{\abovecaptionskip}{-0.1cm}   %调整图片标题与图距离
\setlength{\belowcaptionskip}{1cm}   %调整图片标题与下文距离
\centering
\includegraphics[width=3.5in]{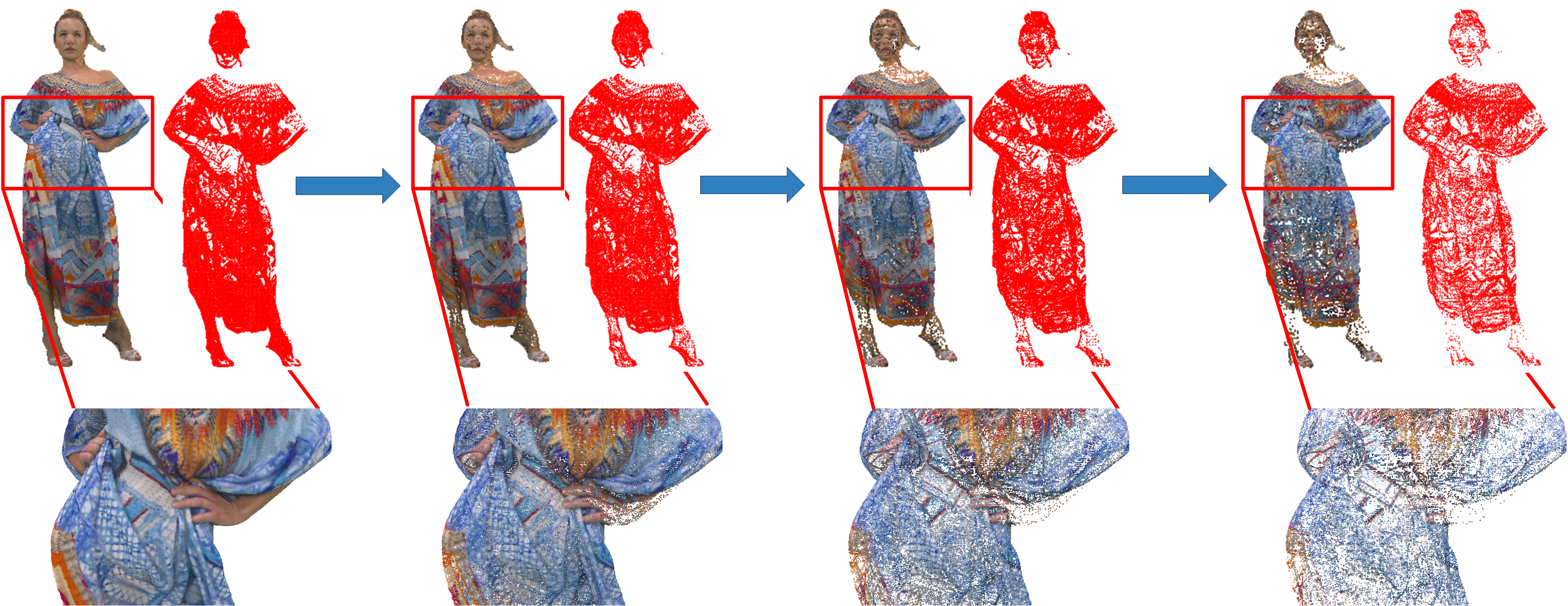}
\caption{Three-stage sampling using the FS module. Each subplot represents the corresponding down sampled point cloud, where the points depicted in red illustrate high-frequency components. Each stage progressively reduces the low-frequency information while focusing on retaining high-frequency components, ensuring better preservation of high-frequency details during compression.}
\label{fig_4}
\end{figure}

\iffalse 
\begin{figure}[!t]
\centering
\includegraphics[width=3.5in]{offatten.png}
\caption{Offset-attention module.}
\label{fig_7}
\end{figure}
\fi 

\subsection{Adaptive Scale Feature Extraction with Geometric Assistance Module}
This module aims to compactly represent color attributes while leveraging geometric information to enhance encoding performance. According to the network structure depicted in Fig. 1, the input point cloud $\bm{\mathcal{P}}_l$ at layer $l$ first undergoes processing through the FS module to obtain the high-frequency component $\bm{\mathcal{P}}_{l+1}$. Next, we compute the set difference,

\begin{equation}
\label{eq26}
\bm{\mathcal{P}}_{l,\textit{\text{res}}} = \bm{\mathcal{P}}_l \ominus \bm{\mathcal{P}}_{l+1}.
\end{equation}

\noindent Then, as shown in Fig. 5, the residual point cloud $\bm{\mathcal{P}}_{l,\textit{\text{res}}}$ is subjected to octree partitioning,
\begin{equation}
\label{eq27}
\tilde{\bm{\mathcal{P}}}_{l,\textit{\text{res}}} = \textit{\text{Octree}}(\bm{\mathcal{P}}_{l,\textit{\text{res}}}),
\end{equation}

\begin{figure}[!t]
\centering
\includegraphics[width=3.5in]{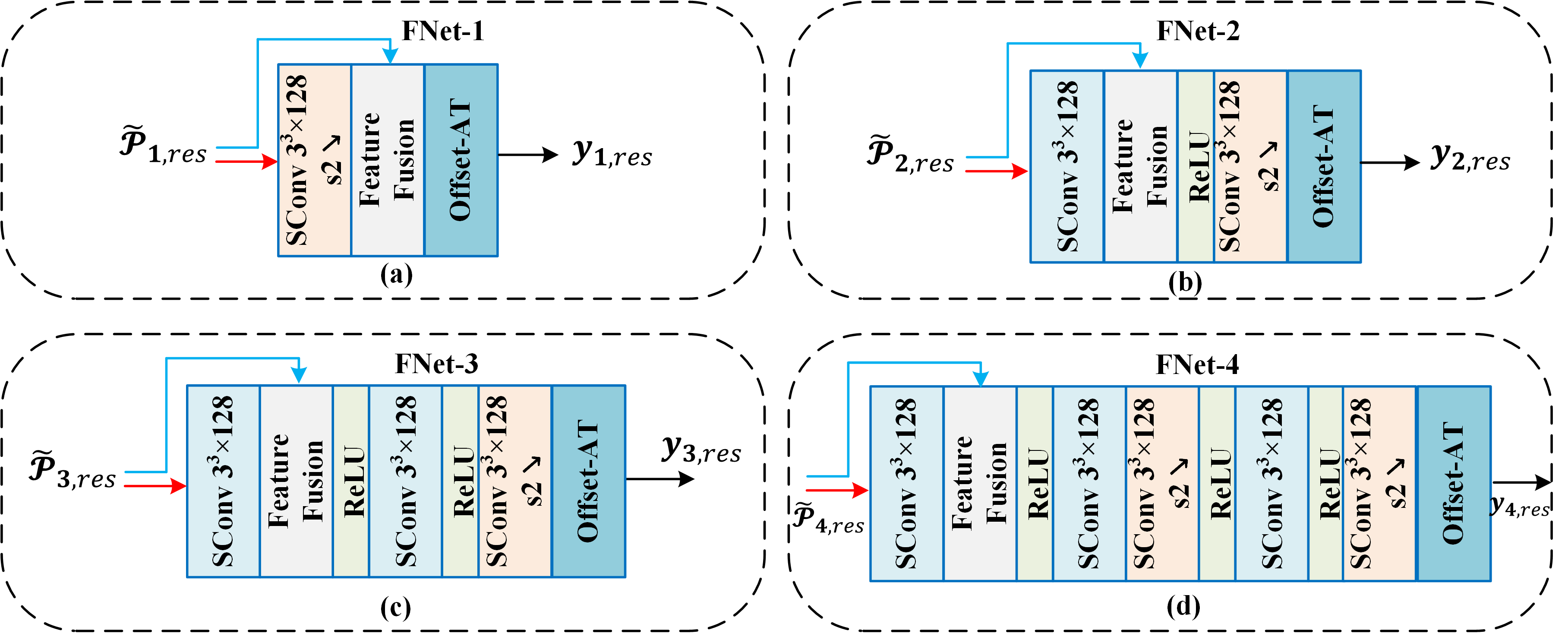}
\caption{Adaptive scale feature extraction modules in the encoding network: (a) FNet-1, (b) FNet-2, (c) FNet-3, and (d) FNet-4. For smooth regions, a shallow feature extraction network, namely FNet-1, is used. For regions with more high-frequency information, a deeper feature extraction network, specifically FNet-4, is used for efficient feature learning and representation.}
\label{fig_6}
\end{figure}

\noindent to recursively divide the 3D space into smaller cubes, organizing the point cloud into a hierarchical structure that facilitates efficient feature extraction and compression. We sort the points by their height (z-axis) in the point cloud, and then partition the point cloud into multiple patches sequentially, based on a fixed number of points. Specifically, each input patch (4096 points) patch  is partitioned to the second-to-last level where each node contains 8 points.

The hierarchical sub-point clouds are then passed through an adaptive scale feature extraction network (FNet) at each layer $l$, as shown in Fig. 4. As the frequency information of different layers varies, the feature extraction networks in each layer should also be different. For smooth components in the upper layers, a shallower feature extraction network is used, while for components with high frequency information in the bottom layers, a deeper feature extraction network is employed for efficient feature learning and representation. FNet integrates convolutional layers and offset-attention \cite{ref58} to capture both local and global features. The convolutional layers extract spatial features, while the offset-attention mechanism enhances the feature representation by capturing dependencies across different parts of the input.

\begin{figure*}[!t]
\setlength{\abovecaptionskip}{-0.1cm}
\centering
\includegraphics[width=5.5in]{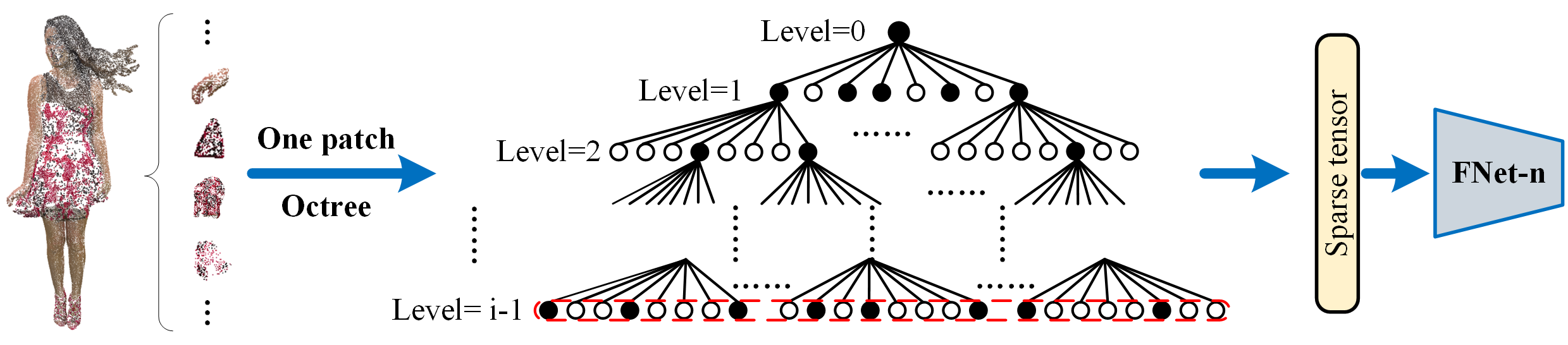}
\caption{Octree partitioning for $\bm{\mathcal{P}}_{l,\text{res}}$ or $\bm{\mathcal{P}}_L$, which is first divided into patches containing 4096 points (except for the last patch). Each patch is then partitioned using an octree to the second-to-last level (containing 8 points per sub-node). After that, the sub-nodes are represented as sparse tensors and fed into the sparse convolution-based feature extraction network.}
\label{fig_5}
\end{figure*}

Because neighbouring points in a local surface often have similar color attributes \cite{ref52}, to better capture the correlation between points, FNet also includes a geometry-assisted attribute feature refinement block in which normal information is calculated based on the geometry and then concatenated with attribute feature for refinement, as shown in Fig. 6. 
\begin{figure}[!t]
\setlength{\abovecaptionskip}{-0.1cm}
\centering
\includegraphics[width=2.5in]{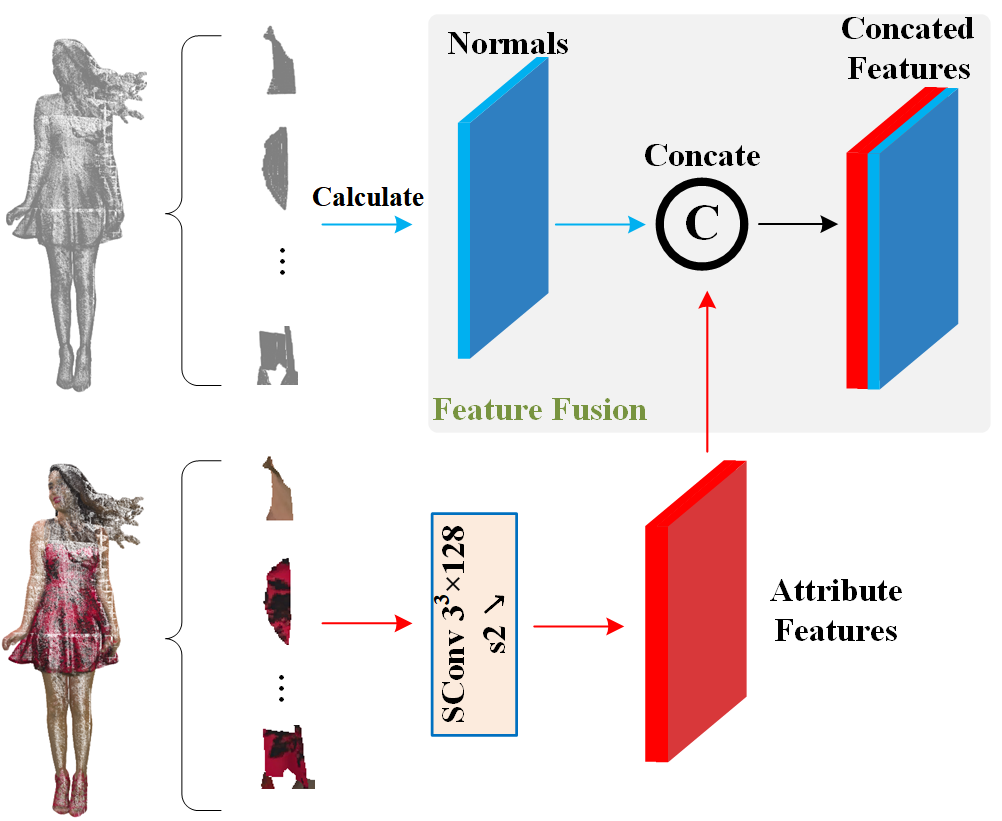}
\caption{Feature Fusion Module. Normals are calculated from the geometric information and concatenated with the attribute features to form new features. These new features are then processed through sparse convolution layers.}
\label{fig_8}
\end{figure}

\subsection{Entropy Encoding and Decoding}
The extracted features at layer $l$, denoted as $\bm{y}_l$, are then quantized and entropy-encoded using a popular hyper model and a specific entropy model, as shown in Fig. 7. The hyper model includes a hyper encoder, a hyper decoder, and a context model, as shown in Fig. 8 (a), (b), and (c), respectively. Specifically, the hyper model uses the feature of the bottom layer (layer $L=4$ in Fig. 1), i.e., $\bm{y}_L$, to generate a global hyperprior information, i.e., $\bm{z}_L$, which is then used to learn a layer-adaptive mean $\mu_l$ and a layer-adaptive scale parameter $\sigma_l$ of a Laplace distribution for efficient entropy encoding and decoding of all the other layers, i.e., $\bm{y}_l, l \in \{1, \ldots, L-1\}$, based on an autoregressive context model which exploits the statistical dependencies between neighbouring elements in the latent space to predict the current element's distribution more accurately. Moreover, a Gaussian distribution-based factorized entropy model \cite{ref59} is leveraged to encode the quantized priors, i.e., $\tilde{\bm{z}}_l$.

In the hyper model, we additionally introduce a scaling noise generation network named HSQ \cite{ref60}, as shown in Fig. 8 (d), as an adaptive quantization operation. This operation implements the quantization adaptively by using additive uniform noise with scaling. First, HSQ generates an adaptive noise scaling factor $\Delta_l$ for each latent element. During training, additive uniform noise $u \sim U\left(-\frac{2}{\Delta_l}, \frac{2}{\Delta_l}\right)$ is added to the latent representation $\bm{y}_l$ to get $\tilde{\bm{y}}_l^{\text{train}}$. During testing, hard quantization is performed according to the generated noise scaling factors, i.e., $\tilde{\bm{y}}_l^{\text{test}} = \Delta_l \cdot \left\lfloor \frac{\bm{y}_l}{\Delta_l} \right\rceil$. The advantage of this adaptive quantization module lies in its ability to dynamically adjust quantization granularity according to the characteristics of each latent variable, thereby enhancing compression efficiency.

\subsection{Decoder}

In the decoding pipeline, the latent features of each layer are first processed by the corresponding ReFNet (Fig. 9). ReFNet transforms the input latent feature $\hat{\bm{y}}_{l,\text{res}}$ of each layer into an initially reconstructed feature $\hat{\bm{y}}_{l,\text{pre}}$:
\begin{equation}
\label{eq28}
\hat{\bm{y}}_{l,\text{pre}} = \textit{\text{ReFNet}}(\hat{\bm{y}}_{l,\text{res}}).
\end{equation}

Next, $\hat{\bm{y}}_{l,\text{pre}}$ is fed into ReconNet, as shown in Fig. 10, to recover the attribute information of each layer by using convolutions and offset-attention:
\begin{equation}
\label{eq29}
\hat{\bm{y}}_{l,\textit{\text{attri}}} = \textit{\text{ReconNet}}(\hat{\bm{y}}_{l,\textit{\text{pre}}}).
\end{equation}

\begin{figure}[!t]
\setlength{\abovecaptionskip}{-0.1cm}
\centering
\includegraphics[width=3.3in]{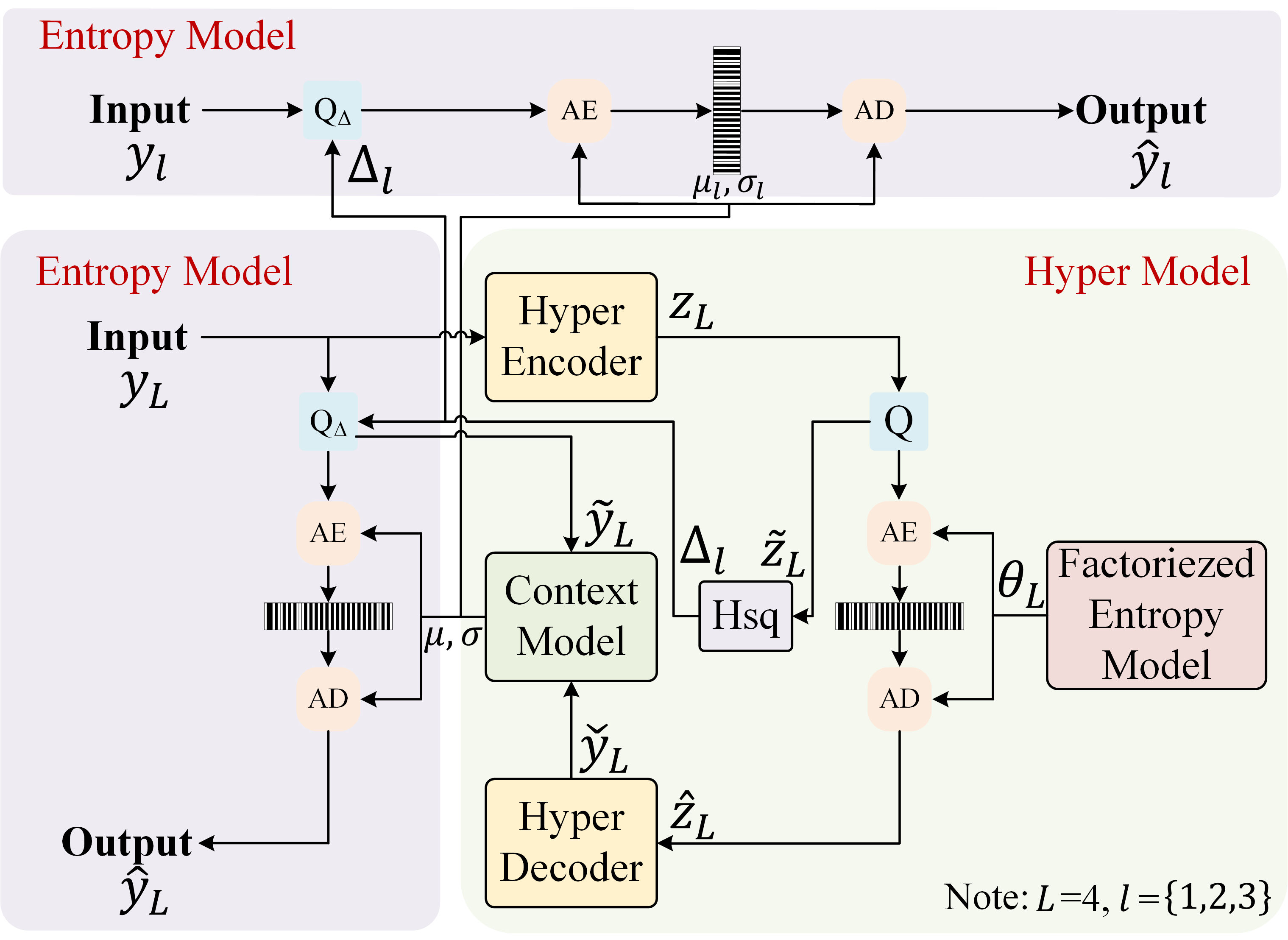}
\caption{Entropy model and hyper model. The hyper model uses the feature of the bottom layer (layer $L=4$), i.e., $\bm{y}_L$, to generate a global hyperprior information, i.e., $\bm{z}_L$, which is then used to learn a layer-adaptive mean $\mu_l$ and a layer-adaptive scale parameter $\sigma_l$ of a Laplace distribution for efficient entropy encoding and decoding of all the other layers, i.e., $\bm{y}_l, l \in \{1, \ldots, L-1\}$, based on an autoregressive context model.
}
\label{fig_9}
\end{figure}

\begin{figure}[!t]
\setlength{\abovecaptionskip}{-0.1cm}
\centering
\includegraphics[width=3.5in]{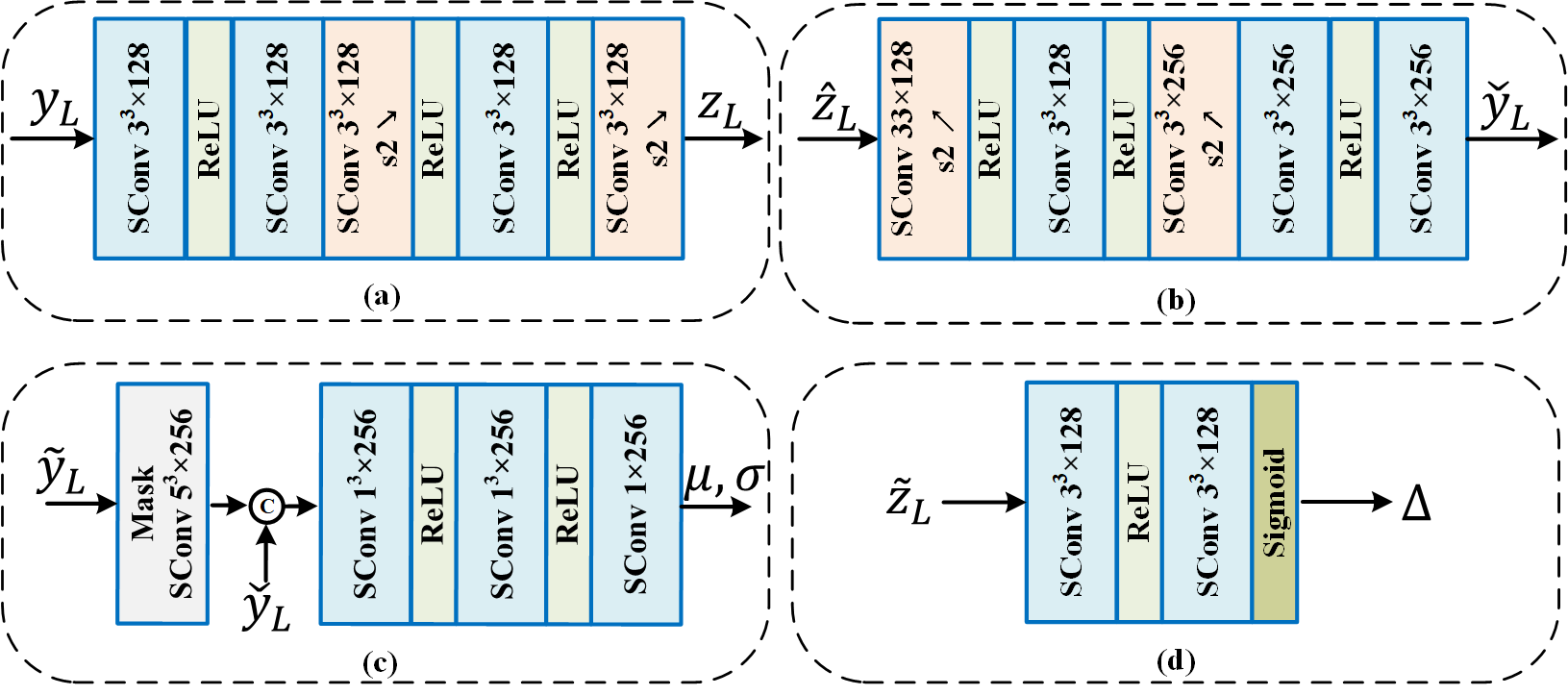}
\caption{ Modules in the hyper model. (a) Hyper encoder, (b) hyper decoder, (c) context model, and (d) HSQ.}
\label{fig_10}
\end{figure}

Throughout the decoding pipeline, the feature of all layers is decoded and reconstructed step by step, forming the complete point cloud attribute. After reconstructing the attribute information of each layer, the known geometric information is combined with it to produce the final reconstructed residual point cloud $\hat{\bm{\mathcal{P}}}_{l,\textit{\text{res}}}$ which is then merged with the bottom layer's reconstructed point cloud $\hat{\bm{\mathcal{P}}}_4$ to progressively restore the reconstructed point cloud:
\begin{equation}
\label{eq30}
\hat{\mathit{\boldsymbol{\mathcal{P}}}}_l = \hat{\bm{\mathcal{P}}}_4 + \sum_{i=1}^l \hat{\bm{\mathcal{P}}}_{i,\textit{\text{res}}}.
\end{equation}
\begin{figure}[!t]
\setlength{\abovecaptionskip}{-0.1cm}
\centering
\includegraphics[width=3.5in]{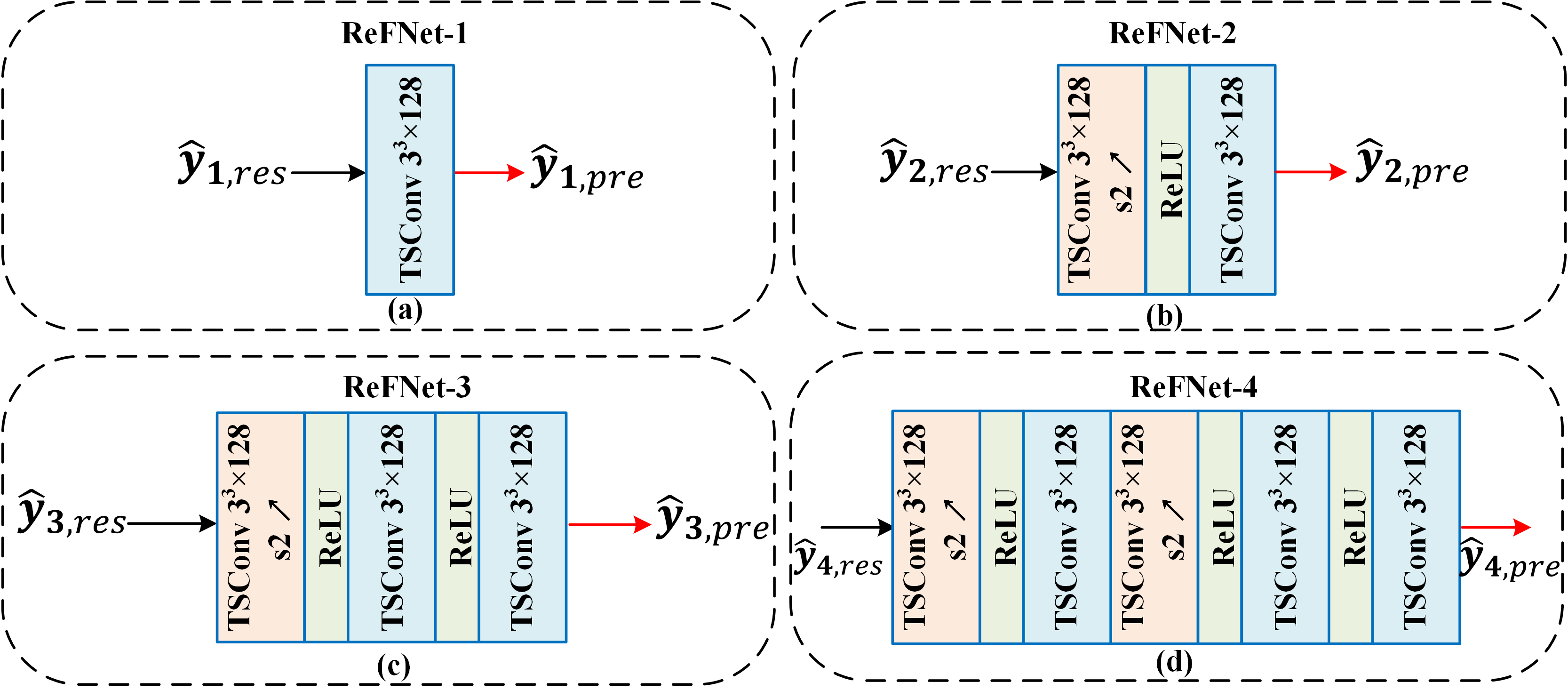}
\caption{ Modules in the decoding network: (a) ReFNet-1, (b) ReFNet-2, (c) ReFNet-3, and (d) ReFNet-4.}
\label{fig_11}
\end{figure}
\begin{figure}[!t]
\setlength{\abovecaptionskip}{-0.1cm}
\centering
\includegraphics[width=2.5in]{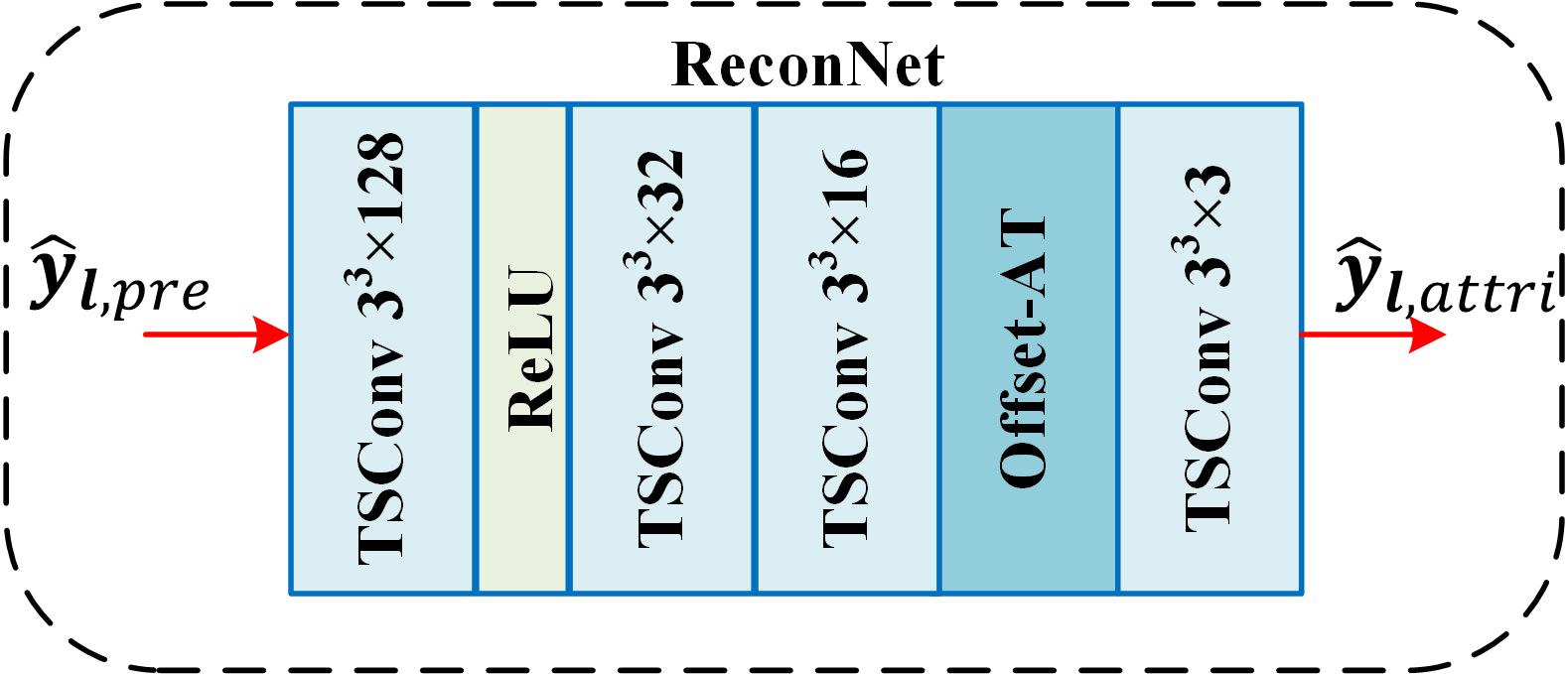}
\caption{ ReconNet.}
\label{fig_12}
\end{figure}

\subsection{Loss Functions}
According to (10), the overall objective is to minimize the distortion under a given bitrate constraint. The reconstruction loss measures the difference between the original attributes and the reconstructed attributes, ensuring that the reconstructed attributes are as close as possible to the original ones. Our loss function is a combination of three losses. The first one is
\begin{equation}
\label{eq31}
D = \sum_{l=1}^L \frac{1}{M_l} \sum_{i=1}^{M_l} \left\| \bm{C}_i^{(l)} - \hat{\bm{C}}_i^{(l)} \right\|^2,
\end{equation}

\noindent where $\textbf{C}_i^{(l)}$ is the color attribute of the input to the $l$th layer, $\hat{\bm{C}}_i^{(l)}$ is the reconstructed color attribute of $\textbf{C}_i^{(l)}$, $L$ is the total number of layers, and $M_l$ is the total number of points in the point cloud. The second loss is the entropy loss based on the probability density function of a Laplace distribution $p_{\mu, \sigma} (\bm{y}_l)$. It ensures efficient compression by minimizing the entropy of the encoded features:
\begin{equation}
\label{eq32}
L_{\text{entropy}} = \sum_{l=1}^L \mathbb{E}_{T_l} \left[ -\log p_{\mu, \sigma} (\bm{y}_l) \right].
\end{equation}

The third loss is a hyperprior loss that captures the discrepancy in hyperprior information, ensuring that the global hyperprior entropy model effectively guides the entropy model. This loss helps in refining the probability distribution parameters $\mu_l$ and $\sigma_l$ used in entropy coding:
\begin{equation}
\begin{aligned}
\label{eq33}
L_{\text{hyperprior}} = & \sum_{i=1}^{N_L} \left[-\log \left( p_{\bm{y}_L^{(i)} | \bm{z}_L^{(i)}} \left( \bm{y}_L^{(i)} | \bm{z}_L^{(i)} \right) \right)\right] + \\
& \sum_{i=1}^{N_L} \left[-\log \left( p_{\bm{z}_L^{(i)}} \left( \bm{z}_L^{(i)} | \bm{\theta}_L \right) \right)\right],
\end{aligned}
\end{equation}

\noindent where $N_L$ represents the number of layers, $\bm{\theta}_L$ represents the set of parameters of the factorized entropy model, $p_{\bm{y}_L^{(i)} | \bm{z}_L^{(i)}} \left( \bm{y}_L^{(i)} | \bm{z}_L^{(i)} \right)$ is the conditional probability model of latent features $y_L$ given hyperpriors $z_L$ in the last layer, and $p_{\bm{z}_L^{(i)}} \left( \bm{z}_L^{(i)} | \bm{\theta}_L \right)$ is the factorized probability model of hyperpriors.

To balance distortion and bitrate, we use an RD loss that combines the reconstruction loss, the entropy loss and the hyperprior loss using Lagrangian multipliers $\lambda_1$ and $\lambda_2$,

\begin{equation}
\label{eq34}
L_{\textit{\text{total}}} = D + \lambda_1 L_{\textit{\text{entropy}}} + \lambda_2 L_{\textit{\text{hyperprior}}}. 
\end{equation}

\textbf{Hybrid approach:} In the proposed method, the encoding of the base layer is interchangeable, allowing for flexible application of traditional methods and neural network-based methods for the other layers. For example, one could use G-PCC to encode and decode the base layer. The global hyperprior entropy model is shifted to the 3rd layer ($l=3$) to leverage hyperprior information from this layer to guide the entropy coding of the other layers. This approach better exploits feature information at different layers and improves coding efficiency (see Section V-D).

\section{Experimental Results and Analysis}
The proposed method was implemented using the PyTorch library, together with the Minkowski Engine, an Intel\textsuperscript{\textregistered} Xeon\textsuperscript{TM} Gold6148 CPU with a base frequency of 2.40 GHz, 32 GB of RAM, and an NVIDIA GeForce RTX 4090 GPU. In Section V-A,, we introduce the detailed configurations of training and testing, In Section V-B, we describe the metrics used for evaluation. In Section V-C, we compare our method with the latest G-PCC test model (TMC13v23) on the MPEG Category Solid and Category Dense datasets. In Section V-D, we compare it with 3DAC \cite{ref49} and ScalablePCAC \cite{ref50}, two state-of-the-art deep learning-based point cloud attribute compression methods. In Section V-E, we illustrate the scalability property of our method. In Section V-F, we compare the time complexity of our method to the benchmark state-of-the-art methods. In Section V-F, we present an ablation study to verify the effectiveness of the different components of our method: FS, number of layers, FNet, geometry-assisted attribute feature refinement, and the global hyperprior entropy model with HSQ.
\iffalse
\begin{figure}[!t]
\centering
\includegraphics[width=0.4\textwidth]{trainingdataset.png}
\caption{ Samples of the training dataset. (a) Shapenet with COCO, (b) Stanford3dDataset and (c) RealWorldTexturedThings}
\label{fig_13}
\end{figure}
\begin{figure}[!t]
\centering
\includegraphics[width=0.4\textwidth]{test dataset.png}
\caption{ Samples of the testing dataset. (a) 8iVFB, (b) MVUB and (c) Owlii}
\label{fig_14}
\end{figure}
\fi 
\subsection{Training and Testing}
\textbf{Training Dataset.} As in \cite{ref34}, we constructed the training set by using the ShapeNet \cite{ref61} and COCO \cite{ref62} datasets. Specifically, we conducted dense sampling on the vertices of ShapeNet meshes, applied random rotations, and quantized the coordinates into 8-bit integers to obtain the coordinates of point clouds. Additionally, we randomly selected images from the COCO dataset and projected them onto the point clouds to assign color attributes. In this way, we generated 15,000 point clouds. To prevent overfitting, we further incorporated the Stanford3dDataset \cite{ref63} into the training dataset.

\begin{table}
\caption{BD-BR(\%) and BD-PSNR (dB) of the proposed method (test codec) vs. G-PCC TMC13v23 (reference codec)\label{tab:table1}}
\centering
\renewcommand{\arraystretch}{1.2}
\resizebox{\linewidth}{!}{%
\begin{threeparttable}
\begin{tabular}{c|c|cc|cc} 
\toprule
\multirow{3}{*}{Class} & \multirow{3}{*}{Point Cloud} & \multicolumn{4}{c}{SPAC vs.G-PCC TMC13v23} \\ 
\cline{3-6}
& & \multicolumn{2}{c|}{BD-BR(\%)} & \multicolumn{2}{c}{BD-PSNR(dB)} \\ 
\cline{3-6}
& & Y & YUV & Y & YUV \\ 
\hline
\multirow{10}{*}{\begin{tabular}[c]{@{}c@{}}Category \\Solid\end{tabular}} & Basketball\_player\_vox11\_0000020 & -21.81 & -18.25 & 0.71 & 0.67 \\
& Dancer\_vox11\_00000001 & -26.21 & -21.24 & 0.79 & 0.66 \\
& Façade\_00064 \_vox11 & -26.91 & -22.13 & 0.64 & 0.69 \\
& Longdress\_vox10\_1300 & -24.75 & -23.75 & 0.65 & 0.59 \\
& Loot\_vox10\_1200 & -25.52 & -23.31 & 0.53 & 0.54 \\
& Queen\_0200 & -23.53 & -21.29 & 0.76 & 0.72 \\
& Redandblack\_vox10\_1550 & -18.97 & -16.82 & 0.59 & 0.51 \\
& Soldier\_vox10\_0690 & -24.55 & -21.17 & 0.71 & 0.66 \\
& Thaidancer\_viewdep\_vox12 & -28.93 & -23.15 & 0.67 & 0.68 \\ 
\cline{2-6}
& \textbf{Average} & \textbf{-24.58} & \textbf{-21.23} & \textbf{0.67} & \textbf{0.63} \\ 
\hline
\multirow{13}{*}{\begin{tabular}[c]{@{}c@{}}Category\\~Dense\end{tabular}} & Boxer\_viewdep\_vox12 & -28.54 & -21.74 & 1.21 & 0.84 \\
& Façade\_00009\_vox12 & -12.76 & -4.77 & 0.18 & 0.05 \\
& Façade\_00015\_vox14 & NA & NA & NA & NA \\
& Façade\_00064\_vox14 & NA & NA & NA & NA \\
& Frog\_00067\_vox12 & -18.71 & -8.17 & 0.22 & 0.15 \\
& Head\_00039\_vox12 & -21.45 & -18.47 & 0.54 & 0.51 \\
& House\_without\_roof\_00057\_vox12 & -13.39 & -9.34 & 0.22 & 0.18 \\
& Landscape\_00014\_vox14 & NA & NA & NA & NA \\
& Longdress\_viewdep\_vox12 & -28.41 & -25.46 & 1.13 & 0.92 \\
& Loot\_viewdep\_vox12 & -28.91 & -21.18 & 0.81 & 0.5 \\
& Redandblack\_viewdep\_vox12 & -23.32 & -20.43 & 0.77 & 0.68 \\
& Soldier\_viewdep\_vox12 & -26.81 & -25.18 & 0.79 & 0.58 \\ 
\cline{2-6}
& \textbf{Average} & \textbf{-22.48} & \textbf{-17.19} & \textbf{0.65} & \textbf{0.49} \\ 
\bottomrule
\end{tabular}
\vspace{2pt}
\begin{tablenotes}
\item \parbox{\linewidth}{Note: For the Façade\_00015\_vox14, Façade\_00064\_vox14, and Landscape\_00014\_vox14, due to their large scale, the method proposed in this paper cannot be directly applied for compression.}
\end{tablenotes}
\end{threeparttable}
}
\end{table}

\textbf{Training Details.} The loss function used for training is defined in (24). To generate bitstreams with varying bitrates, we trained six different models by adjusting $\lambda_1$ in (24). Higher $\lambda_1$ favours lower bitrate at the expense of increased distortion, and vice versa. We set the value of $\lambda_1$ to 1000, 800, 600, 400, 200, and 100 while setting the value of $\lambda_2$ to 1 to train the models. The learning rate was set at 0.0001 and was halved every 500 epochs. The batch size was set to 32, and the models underwent 1,000,000 iterations of training. The Adam optimizer was used for optimization.

\textbf{Testing Dataset.} To validate the efficiency of the proposed method, we conducted tests on five distinct datasets. The first dataset, MVUB \cite{ref64}, encompasses point clouds such as \textit{“Sarah,” “Ricardo,” “Phil,” “David,” and “Andrew”}. The second dataset, 8iVFB \cite{ref65}, includes dynamic human point cloud sequences like \textit{“Longdress,” “Soldier,” “Loot,” and “Redandblack”}. The third dataset, Owlii \cite{ref66}, includes point cloud sequences \textit{“Dancer,” “Exercise,” “Model,” and “BasketballPlayer”}. The last two datasets are categorized as Solid and Dense datasets by MPEG. Category Solid contains point clouds with well-defined shapes, and Category Dense contains point clouds with rich geometric details and complex large-scale scenes. This diverse testing ensures a robust assessment of the performance across various types of content.

\subsection{Performance Evaluation}

We used Bj{\o}ntegaard delta peak signal-to-noise ratio (BD-PSNR) and Bj{\o}ntegaard delta bitrate (BD-BR) \cite{ref67} metrics for RD performance comparison. BD-BR measures the average change in bitrate between two encoding configurations at the same objective quality. The smaller the negative value of BD-BR, the greater the improvement in compression efficiency. BD-PSNR measures the average difference in PSNR between two encoding configurations at the same bitrate or compression level. The larger the BD-PSNR, the greater the improvement in quality. We also compared the RD curves to illustrate the performance qualitatively. The evaluation targeted both the Y component and the YUV composition (we used a 6:1:1 ratio of Y, U, and V components for calculation), offering a comprehensive assessment of the encoding efficiency and quality preservation across different color spaces.

\subsection{Comparison with G-PCC}
\begin{figure}[!t]
\setlength{\abovecaptionskip}{-0.1cm}
\centering
\includegraphics[width=2.5in]{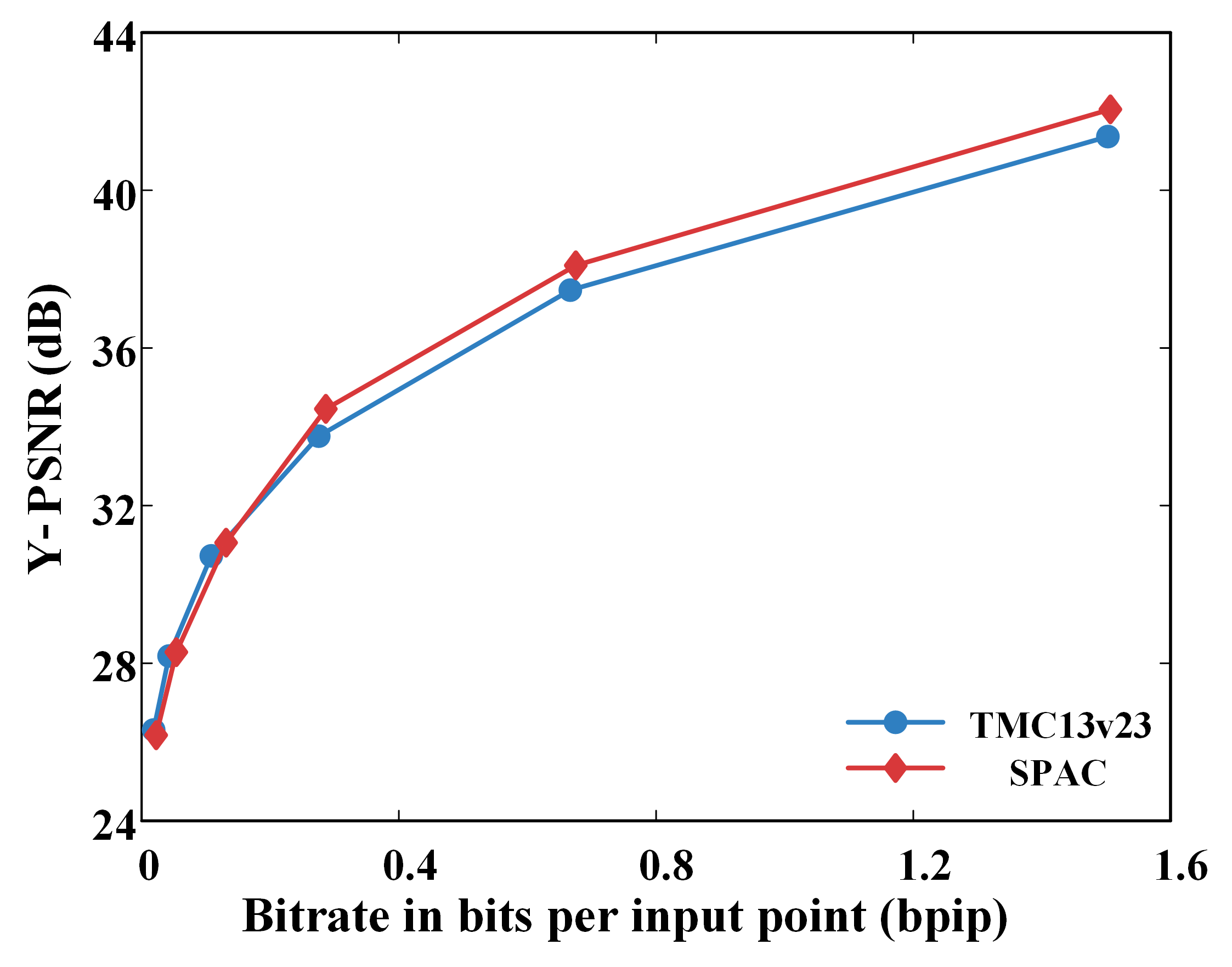}
\caption{Average Y-PSNR vs. average bitrate on categories Solid and Dense under CTCs.}
\label{fig_15}
\end{figure}

We first compared the attribute compression performance of the proposed SPAC with the latest G-PCC test model (TMC13v23), under the RAHT configuration, as shown in Table I and Fig. 11. We can see that SPAC decreased the BD-BR of G-PCC TMC13v23 by 24.58\% (resp. 21.23\%) on Category Solid and 22.48\% (resp. 17.19\%) BD-BR on Category Dense in the Y component (resp. YUV), respectively. Previous works on learning-based attribute compression, such as \cite{ref43,ref45,ref46,ref47,ref49,ref50,ref68}, compared their methods with earlier versions of the G-PCC test model, such as TMC13v6 or TMC13v14, whose compression performance lags behind TMC13v23 significantly, did not strictly follow the MPEG CTCs, or used a very limited subset of the MPEG test datasets. Therefore, the proposed method is the first to significantly surpass G-PCC TMC13v23 under the MPEG CTCs on two large MPEG datasets.

\subsection{Comparison with learning-based methods}
To comprehensively evaluate the effectiveness of the proposed method, we compared it not only with G-PCC but also with the current state-of-the-art learning-based point cloud attribute compression methods, i.e., 3DAC and ScalablePCAC. The test sets included the previously mentioned 8IVFB, Owlii, and MVUB, aligning with their testing methods. As we were unable to obtain the source code of ScalablePCAC, we only compiled the test data based on the published results in the paper \cite{ref50}. The results are summarized in Table II. The proposed method outperformed 3DAC significantly and was comparable with ScalablePCAC. Note that paper\cite{ref50} only provides the coding results of eight point clouds. Note also that ScalablePCAC uses G-PCC to encode its base layer. For a fair comparison, we modified our network to use G-PCC for encoding the last layer (base layer).

\renewcommand{\thefootnote}{\fnsymbol{footnote}}
\begin{table}
\caption{BD-BRs(\%) and BD-PSNRs (dB) of the proposed method (test codec) vs. ScalablePCAC and 3DAC (reference codecs)\protect\footnotemark[1]}
\centering
\renewcommand{\arraystretch}{1}
\resizebox{\linewidth}{!}{%
\begin{threeparttable}
\Large
\begin{tabular}{c|c|cc|cc|cc|cc} 
\toprule
\multirow{3}{*}{Class} & \multirow{3}{*}{Point Cloud} & \multicolumn{4}{c|}{SPAC vs.ScalablePCAC} & \multicolumn{4}{c}{SPAC vs.3DAC} \\ 
\cline{3-10}
& & \multicolumn{2}{c|}{BD-BR(\%)} & \multicolumn{2}{c|}{BD-PSNR(dB)} & \multicolumn{2}{c|}{BD-BR(\%)} & \multicolumn{2}{c}{BD-PSNR(dB)} \\ 
\cline{3-10}
& & Y & YUV & Y & YUV & Y & YUV & Y & YUV \\ 
\hline
\multirow{4}{*}{\begin{tabular}[c]{@{}c@{}}MPEG\\8iVFB\end{tabular}} & Longdress & -0.41 & -0.24 & NA & NA & -36.92 & -36.54 & 1.37 & 1.33 \\
& Soldier & 7.93 & 4.13 & NA & NA & -36.23 & -36.37 & 1.18 & 1.21 \\
& Loot & -6.71 & -5.43 & NA & NA & -38.24 & -38.89 & 1.49 & 1.71 \\
& Redblack & 8.83 & 6.19 & NA & NA & -37.94 & -38.38 & 1.42 & 1.52 \\ 
\hline
\multirow{5}{*}{MVUB} & Sarah & NA & NA & NA & NA & -18.55 & -19.16 & 0.97 & 1.02 \\
& Ricardo & NA & NA & NA & NA & -21.24 & -22.02 & 1.04 & 1.08 \\
& Phil & NA & NA & NA & NA & -16.68 & -17.01 & 0.82 & 0.87 \\
& David & NA & NA & NA & NA & -14.42 & -14.77 & 0.76 & 0.77 \\
& Andrew & NA & NA & NA & NA & -17.15 & -18.21 & 0.91 & 0.95 \\ 
\hline
\multirow{4}{*}{\begin{tabular}[c]{@{}c@{}}MPEG\\Owlii\end{tabular}} & Dancer & 3.43 & -0.53 & NA & NA & -36.89 & -37.54 & 1.23 & 1.28 \\
& Exercise & -11.2 & -8.23 & NA & NA & -2122 & -22.82 & 0.96 & 1.01 \\
& Model & -5.66 & -4.54 & NA & NA & -16.97 & -16.25 & 0.91 & 0.88 \\
& Basketball & 8.11 & 7.64 & NA & NA & -34.71 & -39.1 & 1.15 & 1.62 \\ 
\hline
& \textbf{Average} & \textbf{0.54} & \textbf{-0.13} & \textbf{NA} & \textbf{NA} & \textbf{-26.7} & \textbf{-27.47} & \textbf{1.09} & \textbf{1.17} \\ 
\bottomrule
\end{tabular}
\vspace{4pt}
\end{threeparttable}
}
\end{table}
\begin{table}
\caption{BD-BRs(\%) and BD-PSNRs (dB) of the proposed method with G-PCC encoded base layer (test codec) vs. ScalablePCAC (reference codec)\protect\footnotemark[1]}
\centering
\renewcommand{\arraystretch}{1.1}
\resizebox{\linewidth}{!}{%
\tiny
\begin{threeparttable}
\begingroup
\setlength{\arrayrulewidth}{0.1mm}
\begin{tabular}{c|c|cc|cc} 
\hline
\multirow{3}{*}{Class} & \multirow{3}{*}{Point Cloud} & \multicolumn{4}{c}{SPAC vs.ScalablePCAC} \\ 
\cline{3-6}
& & \multicolumn{2}{c|}{BD-BR(\%)} & \multicolumn{2}{c}{BD-PSNR(dB)} \\ 
\cline{3-6}
& & Y & YUV & Y & YUV \\ 
\hline
\multirow{5}{*}{8iVFB} & Longdress & -8.89 & -9.03 & NA & NA \\
& Loot & -18.71 & -18.86 & NA & NA \\
& Redandblack & 5.53 & 3.27 & NA & NA \\
& Soldier & 5.14 & 3.58 & NA & NA \\ 
\cline{2-6}
& \textbf{Average} & -\textbf{4.23} & -\textbf{5.26} & NA & NA \\ 
\hline
\multirow{4}{*}{Owlii} & Basketball player & 13.04 & 7.33 & NA & NA \\
& Dancer & 3.13 & -0.83 & NA & NA \\
& Exercise & -13.14 & -10.51 & NA & NA \\
& Model & -3.62 & -5.21 & NA & NA \\ 
\cline{2-6}
& \textbf{Average} & -\textbf{0.15} & -\textbf{2.3} & NA & NA \\
\hline
\end{tabular}
\endgroup
\end{threeparttable}
}
\end{table}

Table III compares the performance of our hybrid approach, which uses G-PCC to encode the base layer, to ScalablePCAC. This indicates that combining G-PCC with learning-based point cloud compression profits from the strengths of both approaches, resulting in a more efficient and higher-quality point cloud compression solution than each individual approach.

\footnotetext[1]{Note: The data of ScalablePCAC comes from its paper \cite{ref50}. For a fair comparison with ScalablePCAC, the 8IVFB point clouds in this table have a voxel scale of 12, which differs from the scale used in Table I. While our method and 3DAC encoded all frames of each point cloud sequence, the number of encoded frames for ScalablePCAC was not specified.}

\textbf{Subjective Quality Comparison.} To demonstrate the advantages of the proposed SPAC in terms of subjective quality, we compared the decoded point clouds of SPAC with those of TMC13v23 and 3DAC at similar bitrates, as illustrated in Fig. 12. SPAC was significantly more successful in preserving local texture details(see, for example, the black pattern area of the ``\textit{Redandblack}" skirt and the button area of the ``\textit{Loot}" clothing). This demonstrates SPAC's effectiveness in learning the characteristics of point clouds through the varying frequencies of attributes.

\iffalse 
\begin{figure}[!t]
\setlength{\abovecaptionskip}{-0.1cm}
\centering
\includegraphics[width=2.5in]{averageRDwith3DAC.png}
\caption{Rate-PSNR (Y component) curves comparison between the proposed method and 3DAC. As we were unable to obtain the source code of ScalablePCAC, we did not compare its R-D curves.}
\label{fig_16}
\vspace{-0.4cm}
\end{figure}
\fi 

\begin{figure}[!t]
\setlength{\abovecaptionskip}{-0.1cm}
\centering
\includegraphics[width=3.5in]{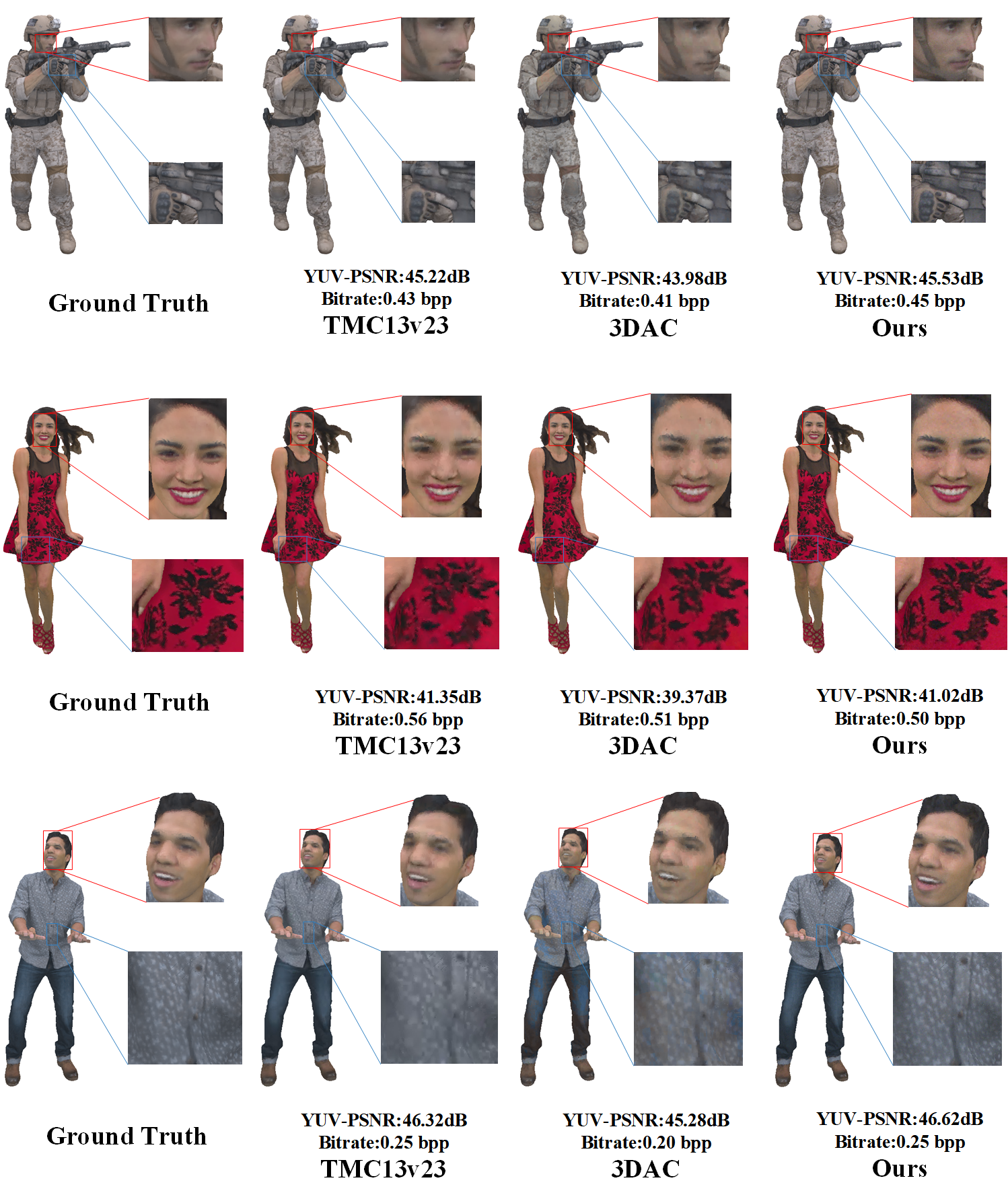}
\caption{ Subjective quality comparison.}
\label{fig_17}
\end{figure}

\begin{figure}[!t]
\setlength{\abovecaptionskip}{-0.1cm}
\includegraphics[width=3.4in]{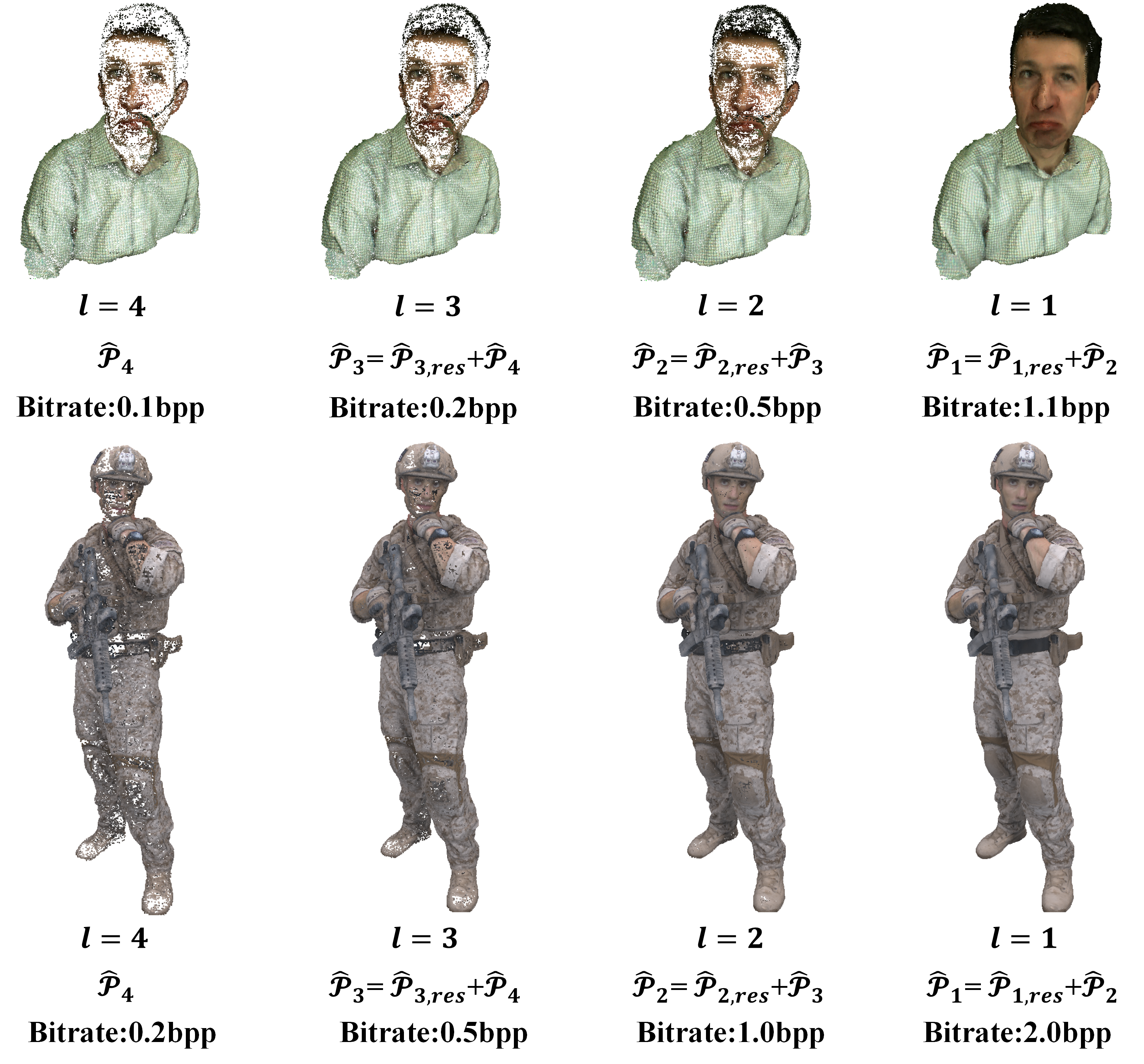}
\caption{Subjective quality for progressive transmission. The reconstruction starts from $l = 4$ (i.e., $\bm{\mathcal{P}}_4$) to preserve the basic structure and high-frequency contours of the point cloud when the bitrate is limited. As the bitrate increases, $\bm{\mathcal{P}}_{3,res}$ can be further reconstructed, and by combining it with $\bm{\mathcal{P}}_4$, $\bm{\mathcal{P}}_3$ can be reconstructed. Furthermore, with additional bitrate, $\bm{\mathcal{P}}_2$ and $\bm{\mathcal{P}}_1$ can also be reconstructed.}
\label{fig_17}
\end{figure}

\subsection{Scalability}
To validate the scalability of the proposed method in point cloud compression, we designed a set of subjective experiments to evaluate the performance of the method at different bitrates. Fig. 13 shows the reconstruction of the point cloud for full-layer compression, three-layer compression, two-layer compression, and base layer compression. When the bitrate is limited, our network encodes the contour information of the point cloud through base layer compression, thereby preserving the overall structure of the point cloud. As the bitrate is gradually increased, our method progressively reconstructs the complete point cloud.

\subsection{Complexity}
Table IV compares the computational complexity of SPAC with G-PCC, 3DAC, and ScalablePCAC. Specifically, we measured the average time for eight point clouds, as shown in Table III. Additionally, G-PCC runs on a CPU, whereas the other methods run on a GPU, making this comparison only a reference. From the results, it can be seen that the encoding and decoding times for SPAC were relatively long compared with G-PCC and 3DAC. This is because we included the sampling time in the encoding stage, which consumed a significant amount of time. The remaining time complexity is due to the autoregressive process in the entropy model, which requires continuous context modeling for each encoded symbol, resulting in high time complexity.
\begin{table}[H]
\caption{Average runtime (seconds per frame)\label{tab:table4}}
\centering
\resizebox{\linewidth}{!}{%
\tiny
\begin{tabular}{c|c|c|c|c} 
\specialrule{0.05em}{0em}{0em}
Method    & SPAC & G-PCC & 3DAC  & ScalablePCAC  \\ 
\hline
Enc. Time (s) & 131.51  & 3.43  & 35.21 & 53,68         \\
Dec. Time (s) & 102.12  & 2.98  & 34.81 & 351.54        \\
\specialrule{0.05em}{0em}{0em}
\end{tabular}
}
\end{table}
\subsection{Ablation Study}
In this sub-section, we conduct a series of ablation studies to understand the contribution of each innovation within the SPAC framework, i.e., FS module, number of layers, adaptive scale feature extraction (FNet) with geometric assistance and the global hyperprior entropy model with HSQ on the coding performance.

\begin{table}[H]
\caption{YUV compound BD-PSNR(db) of SPAC compared to SPAC with FPS\label{tab:table5}}
\centering
\renewcommand{\arraystretch}{1.5}
\resizebox{\linewidth}{!}{%
\Large
\begin{tabular}{c|ccccccc|c} 
\toprule
Point cloud & Longdress & Soldier & RedBlack & Andrew & Ricardo & Sarah & David & Average        \\ 
\hline
BD-PSNR(dB) & 0.41      & 0.35    & 0.17     & 0.26   & 0.18    & 0.2   & 0.28  & \textbf{0.26}  \\
\bottomrule
\end{tabular}
}
\end{table}

\textbf{FS.} We replaced the FS module with a farthest point sampling (FPS), maintaining the same subsequent experimental configuration. The point clouds obtained through FPS had the same number of points as those obtained through the FS module. Table V presents the experimental results on some of the test datasets. We can see that SPAC with FPS led to an average reduction of 0.26dB in BD-PSNR for the YUV composition compared to SPAC with FS.

\begin{table}[H]
\caption{YUV compound BD-PSNR(db) of SPAC compared to SPAC with different number of layers\label{tab:table6}}
\centering
\renewcommand{\arraystretch}{1.5}
\resizebox{\linewidth}{!}{%
\huge
\begin{tabular}{c|c|ccccccc|c} 
\toprule
Number of layers & Point cloud & Longdress & Soldier & RedBlack & Andrew & Ricardo & Sarah & David & Average         \\ 
\hline
2            & BD-PSNR(dB) & 1.65     & 1.04   & 1.51    & 2.03  & 1.33   & 1.42 & 1.28 & \textbf{1.47}  \\ 
\hline
3            & BD-PSNR(dB) & 0.72     & 0.46   & 0.6     & 0.81  & 0.53   & 0.56 & 0.49 & \textbf{0.6}   \\ 
\hline
5            & BD-PSNR(dB) & 0.17     & 0.11   & 0.15    & 0.19  & 0.14   & 0.15 & 0.13 & \textbf{0.15}  \\ 
\hline
6            & BD-PSNR(dB) & 0.41     & 0.34   & 0.39    & 0.5   & 0.35   & 0.37 & 0.35 & \textbf{0.39}  \\
\bottomrule
\end{tabular}
}
\end{table}
\textbf{Number of layers.} The aim of the experiment was to study the effect of the number of layers on encoding efficiency and reconstruction quality. The results, as shown in Table VI, indicate that the proposed four-layer framework led to the best RD performance.

\begin{table}[H]
\caption{YUV compound BD-PSNR(db) of SPAC compared to SPAC  with fixed sparse convolution\label{tab:table7}}
\centering
\renewcommand{\arraystretch}{1.5}
\resizebox{\linewidth}{!}{%
\Large
\begin{tabular}{c|ccccccc|c} 
\toprule
Point cloud & Longdress & Soldier & RedBlack & Andrew & Ricardo & Sarah & David & Average        \\ 
\hline
BD-PSNR(dB) & 1.2       & 1.43    & 1.33     & 1.05   & 0.94    & 1.06  & 1.48  & \textbf{1.21}  \\
\bottomrule
\end{tabular}
}
\end{table}
\textbf{FNet.} To validate the effectiveness of FNets, in this experiment, we replaced FNets and ReFNets with fixed sparse convolution while keeping all other configurations unchanged. Specifically, each layer used a 5-layer sparse convolution with corresponding ReLU operation. Table VII presents the comparative experimental results on some test point clouds. It can be seen that compared to SPAC with FNets, the coding performance of SPAC with fixed sparse convolution significantly decreased, which fully demonstrates that FNets are more effective in extracting features from point clouds of different scales and achieving better performance.

\iffalse 
\begin{table*}[!ht]
    \setlength{\abovecaptionskip}{0.2pt} 
    \caption{BD-BR (\%) for the YUV compound components. The reference codec is SPAC with hyper model and the test codec is SPAC without hyper model.\label{tab:table9}}
    \centering
    \resizebox{14cm}{!}{
    \tiny
    \begin{tabular}{c|ccccccc|c}
    \hline
        Point cloud & Longdress & Soldier & Red\&Black & Andrew & Ricardo & Sarah & David & Average \\ \hline
        BD-BR(\%) & -4.1 & -3.76 & -3.51 & -4.83 & -7.41 & -3.72 & -4.19 & \textbf{-4.5} \\ \hline
    \end{tabular}
    }
    \vspace{-15pt}
\end{table*}
\fi 

\begin{table}[H]
\caption{YUV compound BD-PSNR(db) of SPAC compared to SPAC  without geometric assistance\label{tab:table8}}
\centering
\renewcommand{\arraystretch}{1.5}
\resizebox{\linewidth}{!}{%
\Large
\begin{tabular}{c|ccccccc|c} 
\toprule
Point cloud & Longdress & Soldier & RedBlack & Andrew & Ricardo & Sarah & David & Average        \\ 
\hline
BD-PSNR(dB) & 0.11      & 0.28    & 0.27     & 0.23   & 0.17    & 0.23  & 0.25  & \textbf{0.22}  \\
\bottomrule
\end{tabular}
}
\end{table}
\textbf{Geometric assistance-based feature refinement.} In this experiment, we studied the performance of SPAC with and without the geometry assistance-based feature refinement module. The results in Table VIII show that geometry assistance improved compression efficiency.

\begin{table}[H]
\caption{YUV compound BD-BR(\%) of SPAC compared to SPAC without hypermodel\label{tab:table9}}
\centering
\renewcommand{\arraystretch}{1.5}
\resizebox{\linewidth}{!}{%
\begin{tabular}{c|ccccccc|c} 
\toprule
Point cloud & Longdress & Soldier & RedBlack & Andrew & Ricardo & Sarah & David & Average         \\ 
\hline
BD-BR(\%)   & -4.1      & -3.76   & -3.51    & -4.83  & -7.41   & -3.72 & -4.19 & \textbf{-4.5}  \\
\bottomrule
\end{tabular}
}
\end{table}
\textbf{Global hyperprior entropy model.} To verify the effectiveness of the global hyper prior model, we compared the performance of SPAC with and without it. Table IX highlights the significant improvements in coding efficiency achieved by the global hyper model.

\begin{table}[H]
\caption{YUV compound BD-BR(\%) of SPAC compared to SPAC  without HSQ\label{tab:table8}}
\centering
\renewcommand{\arraystretch}{1.5}
\resizebox{\linewidth}{!}{%
\Large
\begin{tabular}{c|ccccccc|c} 
\toprule
Point cloud & Longdress & Soldier & RedBlack & Andrew & Ricardo & Sarah & David & Average         \\ 
\hline
BD-BR(\%)   & -0.42     & -0.31   & -0.17    & -0.23  & -0.14   & -0.22 & -0.3  & \textbf{-0.26}  \\
\bottomrule
\end{tabular}
}
\end{table}
\textbf{HSQ.} To verify the effectiveness of the adaptive quantization module (HSQ), we conducted an ablation study where we compared the performance of our network with and without this module. Table X shows that the bitrate increased without HSQ, indicating that HSQ was effective in ensuring compression efficiency.

\section{Conclusion}
We presented an end-to-end point cloud attribute coding method that combines a frequency-based sampling network with a multi-layer progressive encoding-decoding structure. Through sampling, octree partitioning, adaptive scale feature extraction, geometry-assisted color feature refinement and a global hyperprior entropy model, our method effciently compresses and reconstructs the color attribute of the input point cloud. Our method exhibits exceptional scalability, making it suitable for applications with various bandwidth and computational resource constraints. The hierarchical structure of the network allows for efficient compression and high-fidelity reconstruction at different layers. Comprehensive experiments and comparisons with the state-of-the-art methods such as G-PCC(TMC13v23), 3DAC, and ScalablePCAC demonstrates that our method achieves significant improvements in both compression efficiency and reconstruction quality. However, our method has two limitations: high computational complexity and inability to process large-scale point clouds. Future work could explore optimizations of the network architecture and hybrid compression strategies to address these issues.

%\newpage

 %needed around the contents of the optional argument to biography to prevent
 %the LaTeX parser from getting confused when it sees the complicated
 %$\backslash${\tt{includegraphics}} command within an optional argument. (You can create
% your own custom macro containing the $\backslash${\tt{includegraphics}} command to make things
%simpler here.)
 

%\bf{If you include a photo:}\vspace{-33pt}
%\begin{IEEEbiography}[{\includegraphics[width=1in,height=1.25in,clip,keepaspectratio]%{fig1}}]{Michael Shell}
%Use $\backslash${\tt{begin\{IEEEbiography\}}} and then for the 1st argument use %$\backslash${\tt{includegraphics}} to declare and link the author photo.
%Use the author name as the 3rd argument followed by the biography text.
%\end{IEEEbiography}


%Use $\backslash${\tt{begin\{IEEEbiographynophoto\}}} and the author name as the argument %followed by the biography text.

\vfill

\end{document}